\documentclass[showpacs,preprintnumbers,amsmath,amssymb,twocolumn,superscriptaddress,prb]{revtex4}
\usepackage{graphicx}
\usepackage{amsfonts}
\usepackage{amsmath, amsthm, amssymb}
\usepackage{dsfont}
\usepackage{times}
\usepackage{verbatim}

\newcommand{\Imag}{{\Im\mathrm{m}}}   
\newcommand{\Real}{{\mathrm{Re}}}   
\newcommand{\talpha}{\tilde{\alpha}}
\newcommand{\ve}[1]{{\boldsymbol{#1}}}
\newcommand{\vk}{\ve{k}} 
\newcommand{\vecr}{\ve{r}} 

\newcommand{\e}[1]{\mathrm{e}^{#1}}
\newcommand{\pdiff}[2]{\frac{\partial #1}{\partial #2}}

\newcommand{\cf}{\textit{cf. }}
\newcommand{\ie}{\textit{i.e. }}
\newcommand{\eg}{\textit{e.g. }}
\newcommand{\etal}{\emph{et al.}}
\def\i{\mathrm{i}}

\begin{document}
\title[Quantum transport in ballistic $s_\pm$-wave superconductors with interband coupling: conductance spectra, 
Josephson current, and crossed Andreev reflection]{Quantum transport in ballistic $s_\pm$-wave superconductors 
with interband coupling: conductance spectra, crossed Andreev reflection, and Josephson current}
\author{Iver Bakken Sperstad}
\affiliation{Department of Physics, Norwegian University of
Science and Technology, N-7491 Trondheim, Norway}
\author{Jacob Linder}
\affiliation{Department of Physics, Norwegian University of
Science and Technology, N-7491 Trondheim, Norway}
\author{Asle Sudb{\o}}
\affiliation{Department of Physics, Norwegian University of
Science and Technology, N-7491 Trondheim, Norway}

\date{Received \today}
\begin{abstract}
\noindent We study quantum transport in ballistic $s_\pm$-wave superconductors where coupling between the two bands 
is included, and apply our model to three possible probes for detecting the internal phase shift of such a pairing 
state: tunneling spectroscopy in a N$|s_\pm$-wave junction, crossed Andreev reflection in a two-lead N$|s_\pm$-wave$|$N 
system, and Josephson current in a $s$-wave$|$I$|s_\pm$-wave Josephson junction. Whereas the first two probes are insensitive 
to the superconducting phase in the absence of interband coupling, the Josephson effect is intrinsically phase-dependent, 
and is moreover shown to be relatively insensitive to the strength of the interband coupling. Focusing on the Josephson 
current, we find a 0-$\pi$ transition as a function of the ratio of effective barrier transparency for the two bands, 
as well as a similar phase-shift effect as a function of temperature. An essential feature of this $s_\pm$-wave model 
is non-sinusoidality of the current-phase relation, and we compute the dependence of the critical current on an 
external magnetic field, showing how this feature may be experimentally observable for this system. We also comment 
on the possible experimental detection of the phase shift effects in $s_\pm$-wave superconductors.
\end{abstract}	
\pacs{74.20.Rp, 74.50.+r, 74.70.Dd}

\maketitle

\section{Introduction}

During the last few years, multiband superconductivity has again been at the forefront of condensed matter physics, and particularly 
so after the discovery of high-temperature superconductivity in the family of intrinsically multiband iron-based 
materials\cite{kamihara,mazin_review,ishida-hosono_review}. As with all newly discovered superconductors with unconventional behavior, one principal 
question is to determine the pairing symmetry of the superconductor. In the pnictide superconductors much effort has been devoted to
this central issue, so far without entirely conclusive answers. Nevertheless, the leading contender has for some time been 
$s_\pm$-wave pairing\cite{mazin-singh-johannes_pairing}, which in its simplest realization for the iron-based superconductors means 
that the hole-like and electron-like Fermi surfaces both host $s$-wave superconductivity, but with opposite sign of the order 
parameter. (In the past, similar sign-shifted order parameters have also been considered as a candidate pairing state \eg of high-T$_c$ cuprates\cite{Golubov-Mazin_spin-ladder}.) 

Distinguishing such a state from an isotropic $s$-wave pairing state is highly non-trivial, since both $s$-wave and 
$s_\pm$-waves states have the same symmetry, and do not have nodes in the order parameter on the Fermi surface. 
In order to establish conclusively the internal phase shift characterizing a possible $s_\pm$-wave state in the iron-based superconductors it is therefore crucial to devise phase sensitive pairing probes. A large number of proposals for such experiments have been put 
forth in the literature recently. Theories for multiband tunneling spectroscopy have been developed\cite{wang_multiband_tunneling,
 nagai_n-band,feng_multiband_nss, golubov-tanaka_multiband_btk, araujo-sacramento_multiband_andreev,linder_iron-based} as well 
as calculations of the surface density of states for a $s_\pm$-superconductor\cite{onari-tanaka_spm_sdos, choi_spm-pairing_bdg_ldos}. 
In a related context, Andreev bound states (ABS) are often pointed out as possible pairing probes 
\cite{ghaemi-wang-vishwanath_abs_probe, tsai_cc-pairing_impurity-induced_ABS,zhou_cc-pairing_impurity-induced_abs, matsumoto_impurity_bound_state, ng_impurity_bound_states}. 
Another class of experiments suggested involves Josephson junctions, both single
junctions\cite{tsai-bernevig_cc,ota_multigap_sis,ota_josephson_vortex,Ota_intergrain,ng_nagaosa}, trijunction
loops\cite{chen_pi-trijunction,chen_trijunction_half_flux-quantum_jumps} and also various 
corner geometries employed for Josephson interferometry\cite{wu_phillips_iron_squid, parker-mazin_phase_sensitive}. 
Yet another work considered possible signatures in the AC Josephson effect\cite{inotani_pm_riedel}. In addition, we should mention that the Josephson effect for multiband superconductors with sign-shifted order parameters has previously been discussed also in the context of MgB$_2$\cite{Agterberg_multiband} and bilayer cuprates\cite{mazin_YBCO_JJ}.

\begin{figure}[h]
	\centering
	\resizebox{0.45\textwidth}{!}{
	\includegraphics{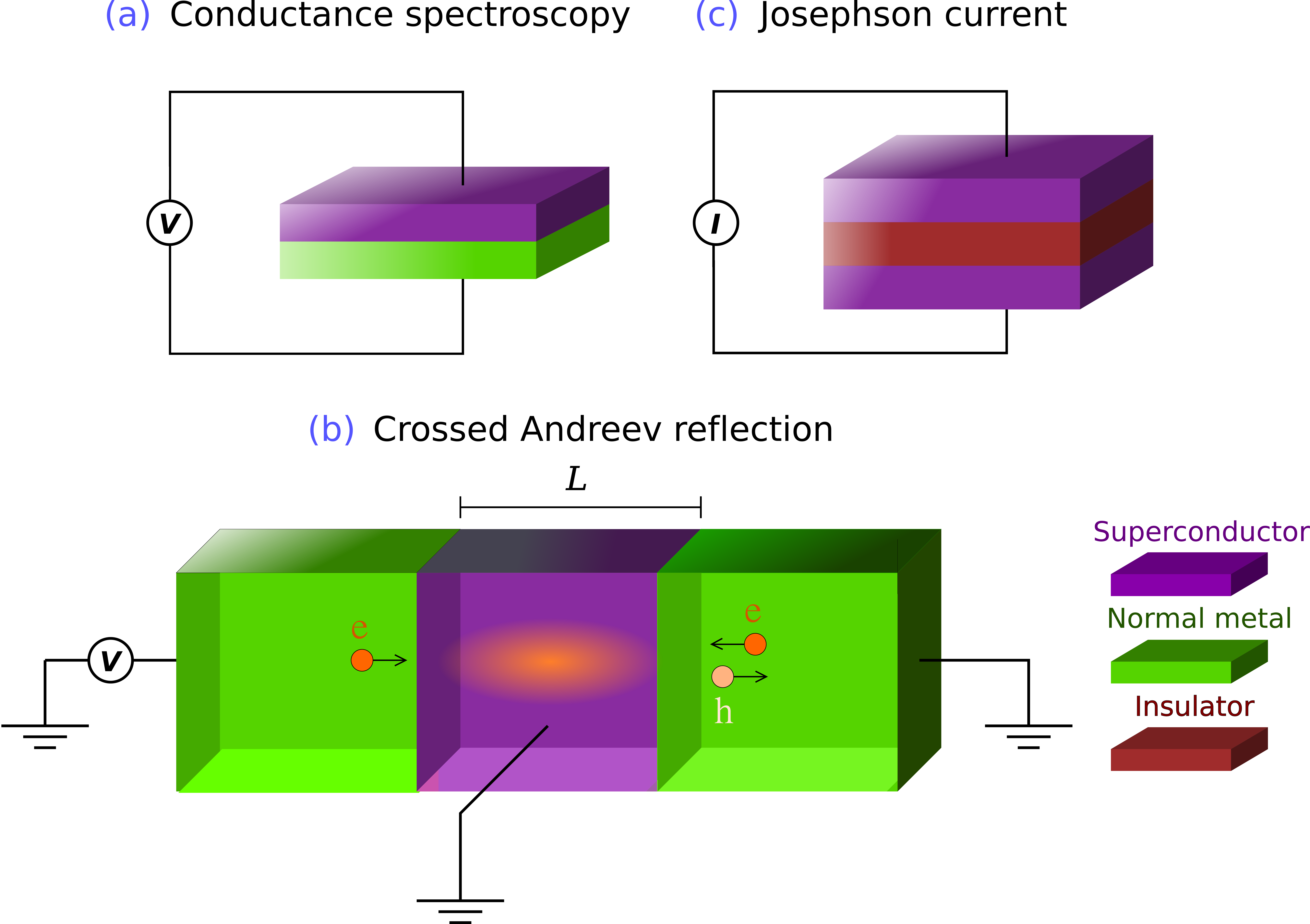}
	} \caption{(Color online) Schematic drawing of the systems under consideration in this work: a) The model of N$|s_\pm$-wave junction for 
	tunneling spectroscopy as studied in Sec. \ref{sec:cond}, b) the model of the two-lead N$|s_\pm$-wave$|$N junction for the study of 
	crossed Andreev reflection in Sec. \ref{sec:car}, and c) the model of the $s$-wave$|$I$|s_\pm$-wave Josephson junction considered 
	in Sec. \ref{sec:jos}. For system (b), we have illustrated how an electron in the left-hand lead is converted to a hole in the 
	right-hand lead by the formation (together with a electron from the right-hand lead) of a Cooper pair in the superconducting 
	interlayer.}
	\label{fig:model}
\end{figure}

Of the probes listed above, tunneling spectroscopy is probably the one that is experimentally most accessible 
(see Refs. ~\onlinecite{chen-tesanovic_review}, ~\onlinecite{gonnelli_2_nodeless_gaps} and references therein), 
and results here are routinely compared with the theory of Blonder, Tinkham and Klapwijk (BTK) for Andreev reflection\cite{btk}. 
Recently, one theoretical work\cite{golubov-tanaka_multiband_btk} augmented the BTK-approach to also incorporate interband 
scattering in the superconducting region, which was shown to result in interference effects and subgap bound states 
in the conductance spectra. However, as pointed out soon after\cite{araujo-sacramento_multiband_andreev}, the phenomenological 
approach employed in Ref. ~\onlinecite{golubov-tanaka_multiband_btk} may fail to capture the effect of interband coupling correctly. 
In this work, we will present an alternative approach of including interband scattering into the BTK framework.

Another probe which has not been considered in the literature so far is crossed Andreev reflection\cite{deutscher_car} (CAR). This is a 
process contributing to the nonlocal conductance in a two-lead normal metal / superconductor junction\cite{byers-flatte_car} in which an 
electron impinging on the superconductor from one of the leads is converted to a hole in the other lead. This phenomenon has previously 
attracted attention as a possible probe both for ferromagnetic superconductors\cite{benjamin_car_fmsc} and non-centrosymmetric 
superconductors \cite{fujimoto_car_noncentro}. However, crossed Andreev reflection has not yet, to the best of our knowledge, been considered in the context of the $s_{\pm}$-wave pairing 
state. 

Yet another possible experimental signature, which was first proposed in the context of iron-based superconductors by the present authors in Ref. ~\onlinecite{linder_iron_rapid}, 
is 0-$\pi$ transitions \cite{ryazanov_pi-junction, buzdin_review}. To explain this phenomenon, we draw upon results from Josephson 
junctions with ferromagnetic elements. For such systems, \eg a S$|$F$|$S junction, the critical current $I_c$ switches sign for 
given thicknesses $d_F$ of the ferromagnetic interlayer, resulting in non-monotonous dependence of $I_c$ on $d_F$. This phenomenon 
is ascribed to the junction switching between being a (conventional) 0-junction with zero phase difference across the junction 
in the ground state and a $\pi$-junction, which has phase difference $\pi$ across the junction in the ground state. 
Furthermore, the critical thicknesses $d_F$ of S$|$F$|$S systems are often temperature dependent, which allows for the observation 
of thermally induced 0-$\pi$ transitions at $T = T_{0 \pi}$ as well as transitions as a function of interlayer width.

The possibility of $\pi$-junctions consisting of $s_\pm$-wave superconductors has been mentioned previously in some theoretical works\cite{chen_pi-trijunction, chen_trijunction_half_flux-quantum_jumps, avishai-ng_multiband_jospehson_anderson_imp, tsai-bernevig_cc}, 
but in Ref. ~\onlinecite{linder_iron_rapid} we showed that 0-$\pi$ transitions were possible in a diffusive $s$-wave$|$N$|s_\pm$-wave junction 
both as a function of temperature and as a function of the ratio of interface resistances for each band. The present work is motivated 
by the question of whether these effects persist in the ballistic limit, and we perform a complementary, more comprehensive study of 
the Josephson effects for a simple model capturing the essential features of a $s_\pm$-wave superconductor with interband coupling. 
We find that the 0-$\pi$ transition for varying ratio of interband resistance persists, but that non-sinusoidality of the current-phase 
relation is significant for the present case. For varying temperature we find a somewhat weaker phase shift effect, which we will 
relate to the more clear-cut 0-$\pi$ transition reported for the diffusive case. These results for the temperature dependence of the Josephson current can be compared with the non-monotonous Josephson current between a multigap and a single-gap superconductor previously obtained by Agterberg \etal\cite{Agterberg_multiband}

The outline of this work is as follows. In Sec. \ref{sec:theory} we present the theoretical framework that is employed to obtain 
our results. This framework will then be applied first to tunneling spectroscopy of a N$|s_\pm$-wave structure in 
Sec. \ref{sec:cond}, after which we will turn to the study of crossed Andreev reflection in a N$|s_\pm$-wave$|$N junction in 
Sec. \ref{sec:car}. The Josephson junction, to which we will devote the largest share of attention, will be treated in 
Sec. \ref{sec:jos}. The three experimental setups are shown schematically in Fig. \ref{fig:model}. Some aspects of our 
model and the possible physical realization of the effects found here are discussed in Sec. \ref{sec:discussion}, and 
we conclude the present work in Sec. \ref{sec:concl}.

\section{Theory}
\label{sec:theory}

We consider the Bogoliubov-de Gennes (BdG) equations for a two-band superconductor with dispersions $\varepsilon_{\vk,\lambda}$ measured 
from the Fermi level $E_{\rm  F}$ and gaps $\Delta_\lambda$, $\lambda=\{1,2\}$, which read
\begin{align}
	\label{eq:ham}
	\begin{pmatrix}
		\hat{H_1} & \hat{0}\\
		\hat{0} & \hat{H_2} \\
	\end{pmatrix}
	\begin{pmatrix}
		\psi_1 \\
		\psi_2 \\
	\end{pmatrix}
	= E \begin{pmatrix}
		\psi_1 \\
		\psi_2 \\
	\end{pmatrix},\;
	\hat{H_\lambda} = \begin{pmatrix}
		\varepsilon_{\vk,\lambda} & \Delta_\lambda \\
		\Delta_\lambda^* & -\varepsilon_{\vk,\lambda} \\
	\end{pmatrix}.
\end{align}
Above, we have used a fermion basis
\begin{equation}
	\label{eq:basis}
	\Psi_\vk = [\eta_{1,\vk}^\dag, \eta_{1,-\vk}, \eta_{2,\vk}^\dag, \eta_{2,-\vk}],
\end{equation}
where $\eta_{\lambda,\vk}$ are fermion operators for band $\lambda$. Considering here positive excitation energies $E>0$, the solution 
for the wavefunctions $\psi_\lambda$ is obtained as a generalized BCS expression:
\begin{align}
	\psi_\lambda &= \Bigg\{\begin{pmatrix}
		u_\lambda\\
		v_\lambda \e{-\i\phi_\lambda}\\
	\end{pmatrix},\; 
	\begin{pmatrix}
		v_\lambda\e{\i\phi_\lambda}\\
		u_\lambda \\
	\end{pmatrix}\Bigg\}, 
\end{align}
where the coherence functions are
\begin{align}
\label{eq:coherence}
	u_\lambda^2 = 1-v_\lambda^2 = \frac{1}{2}\Big(1 + \sqrt{E^2-|\Delta_\lambda|^2} / E \Big),
\end{align}
while the phases $\phi_\lambda$ correspond to the broken U(1) gauge symmetry of the superconducting state. For the $s_\pm$ state, we 
have $\phi_1-\phi_2 = \pi$. Note that in Eq. \eqref{eq:ham}, no assumptions have been made about the pairing mechanism responsible 
for the presence of energy gaps $\Delta_\lambda$ in our model, nor of the origin of a possible internal phase shift. Our 
motivation in this work is merely to investigate the experimental consequences of such a phase shift, when present.

In order to capture interference effects between the bands, it is important to consider carefully the boundary conditions 
in the presence of interband coupling. The above scenario corresponds however to a two-band superconductor with no explicit 
coupling between the bands. (Once again, since we make no assumptions on the pairing mechanism, the gaps of the two bands 
in Eq. \eqref{eq:ham} might be implicitly coupled through two-particle scattering processes, although whether or not this 
would be the case in a microscopic theory will have no consequences for the present model.) Hopping between the bands 
will be taken into account by adding a single-particle hopping term $H_\text{hop}$ to the Hamiltonian:
\begin{align}
	\label{eq:hop}
	H_\text{hop} &= \alpha \int \text{d}\vecr [ \eta_1(\vecr)\eta_2^\dag(\vecr) + \eta_2(\vecr)\eta_1^\dag(\vecr) ],
\end{align}
where $\eta(\vecr)$ are fermion field operators in real space, while $\alpha$ denotes the hopping parameter. Upon including the 
standard delta-function barrier potential $V_0$ at an interface, one may then write down the full BdG-equations in the system. 
Let us, to be definite, consider an N$\mid$$s_\pm$ junction, where the superconductor occupies the halfspace $x>0$. We then 
have $\hat{H}\Psi = E\Psi$, where
\begin{widetext}
\begin{align}
	\label{eq:bdg}
	\hat{H} &= \begin{pmatrix}
		\varepsilon_{\vk,1} + V_0\delta(x) & \Delta_1\Theta(x) & \alpha\delta(x) & 0\\
		\Delta_1^*\Theta(x) & -\varepsilon_{\vk,1} - V_0\delta(x) & 0 & -\alpha\delta(x)\\
		\alpha\delta(x) & 0 & \varepsilon_{\vk,2} + V_0\delta(x) & \Delta_2\Theta(x)\\
		0 & -\alpha\delta(x) & \Delta_2^*\Theta(x) & -\varepsilon_{\vk,2} - V_0\delta(x) \\
	\end{pmatrix}
\end{align}
\end{widetext}
It is seen that the two bands couple through the interface scattering as long as $\alpha\neq0$, in a simple model which nevertheless should 
be able to capture the main qualitative effects.

Before turning to applications of this theory, we state the resulting boundary conditions for the N$\mid$$s_\pm$ junction. For an 
incoming electron from band $\lambda'=1$ on the N side ($x \leq 0$), we write the wavefunction as
\begin{align}
	\label{eq:psiN}
	\psi_N &= [1,0,0,0](\e{\i k x} + r_1\e{-\i kx}) + r^A_1[0,1,0,0] \e{\i kx} \notag\\
	&+ r_2[0,0,1,0]\e{-\i kx} + r^A_2[0,0,0,1] \e{\i kx},
\end{align}
where $k = k_{\rm F}$. Here and in what follows, we assume that the Fermi level $E_{\rm  F}$ is much larger than 
$(\Delta_\lambda,E)$, such that the wavevectors simply read $k_F = \sqrt{2m E_{\rm  F}}$. We also take $E_{\rm F}$ 
to be the same everywhere in the system, since the effect of any Fermi wavevector mismatch (FWVM) can be accounted 
for by adjusting the barrier transparency. Note that although the formalism used in Eq. \eqref{eq:psiN} imposes 
the multiband basis also on the normal metal wavefunction, this does 
not necessarily imply that the normal metal has two physically distinct bands. 

For an incoming electron from band $\lambda'=2$, the 
N side wave function is simply obtained by letting $[1,0,0,0] \e{\i k x}$ go to $[0,0,1,0]\e{\i k x}$ in Eq. \eqref{eq:psiN}.
Here, $\{r_\lambda, r^A_\lambda\}$ are the normal and Andreev reflection scattering coefficients for band $\lambda$. We let the 
wavefunction on the superconducting side ($x > 0$) be unspecified for the moment. The general boundary conditions can then found 
from Eq. \eqref{eq:bdg} as
\begin{align}
	\label{eq:bc}
	\psi_N(x=0) &= \psi_S(x=0),\notag\\
	(\partial_x \psi_S - \partial_x \psi_N) |_{x=0} &= 2m[V_0 \ \mathrm{diag}(\hat{1},\hat{1}) \notag \\
	&+ \alpha \ \mathrm{offdiag}(\hat{1},\hat{1})] \psi_N,
\end{align}
where $\hat{1}$ is the $2 \times 2$ unit matrix and $\mathrm{diag}$ and $\mathrm{offdiag}$ denote diagonal and off-diagonal $4 \times 4$ 
block matrices in which these unit matrices are embedded.
At this point we also introduce two dimensionless parameters characterizing the system, namely the barrier strength $Z = 2 m V_0/k$ 
and the interband coupling strength $\tilde{\alpha} = 2m \alpha / k$.

\section{Results}
\label{sec:results}

\subsection{Conductance spectra}
\label{sec:cond}

As a first application of our model, we calculate the conductance of a N$\mid$$s_\pm$-wave junction and compare it to that 
of its $s$-wave counterpart. This was also done in Ref. ~\onlinecite{golubov-tanaka_multiband_btk}, but in contrast to their 
approach, we construct our wavefunctions and boundary conditions from the full $4\times4$ BdG-equations, as required for a 
multiband scenario. In this case, the wavefunction on the superconducting side reads
\begin{align}
	\label{eq:psiS}
	\psi_S &= s_1[u_1,v_1\e{-\i\phi_1},0,0] \e{\i kx} + t_1[v_1\e{\i\phi_1},u_1,0,0]\e{-\i kx} \notag\\
	&+ s_2[0,0,u_2,v_2\e{-\i\phi_2}] \e{\i kx} + t_2[0,0,v_2\e{\i\phi_2},u_2]\e{-\i kx},
\end{align}
with $\{s_\lambda, t_\lambda\}$ being the transmission coefficients for band $\lambda$. We will use the gauge $\phi_1 = 0$, and 
make explicit use of the internal phase shift by writing $\e{\i(\phi_1-\phi_2)} \equiv \delta = \pm 1$ for the superconductor 
being a two-band $s$-wave superconductor or a $s_\pm$-wave superconductor, respectively. For the normal metal side, we use 
$\psi_N$ from Eq. \eqref{eq:psiN}. We can then solve Eqs. \eqref{eq:bc} for the given wavefunctions, but as the resulting 
expressions for $\{r_\lambda, r^A_\lambda, s_\lambda, t_\lambda\}$ do not allow a simple interpretation in our case, we 
give the solution in Appendix A. 

\begin{figure}[h]
	\centering
	\resizebox{0.45\textwidth}{!}{
	\includegraphics{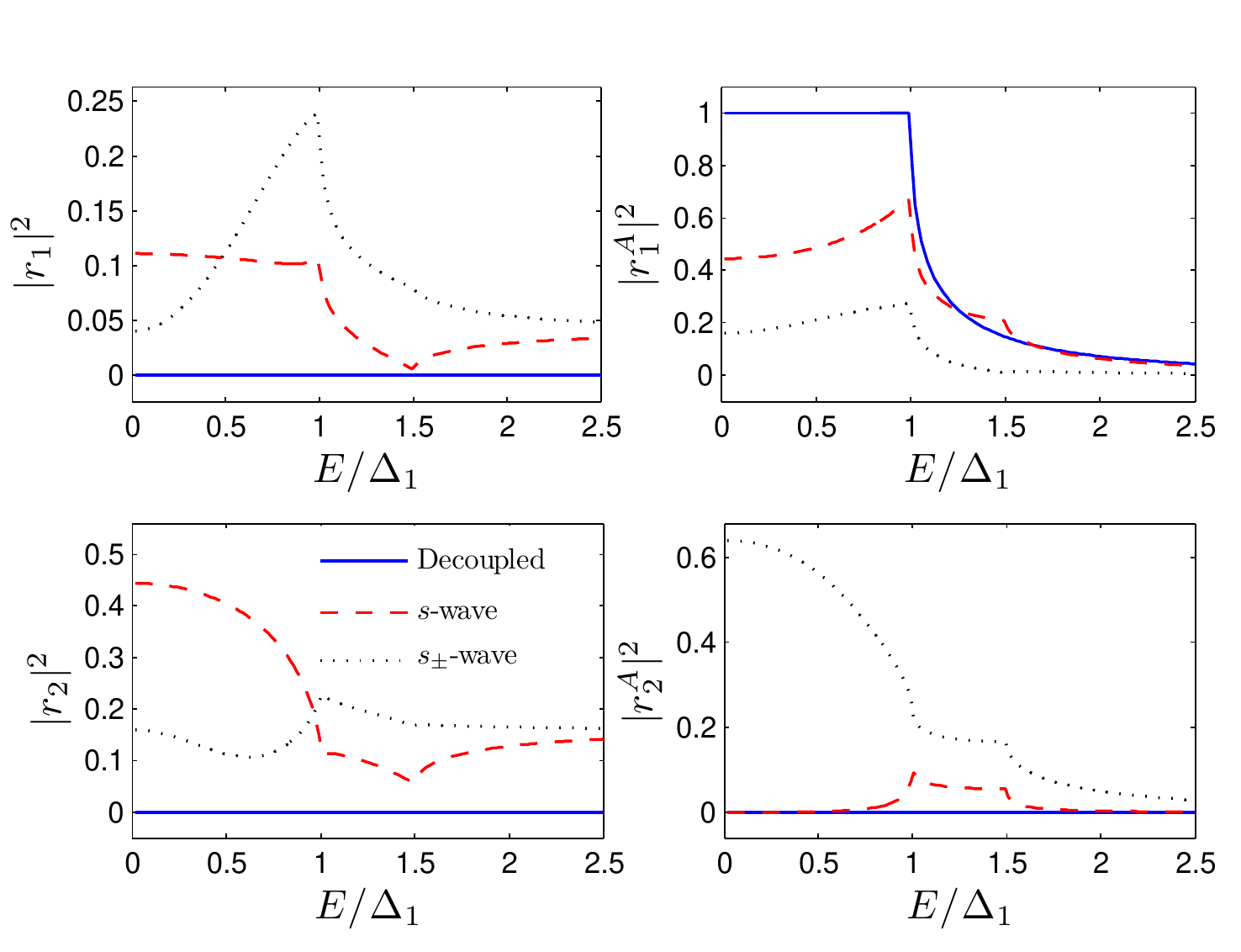}
	} \caption{(Color online) Comparison of the probabilities of the reflection processes in a N$|s_\pm$-wave junction and a two-band 
	N$|s$-wave junction as described in the text. We have chosen zero barrier strength, $Z = 0$, and gap ratio $r_\Delta = 1.5$. We 
	have used a value $\talpha = 1$ for the interband coupling (for the $s$-wave and $s_\pm$-wave case), while for the decoupled case we have $\talpha = 0$.}
	\label{fig:probs_Z0_a05_rD15}
\end{figure}

To illustrate the influence of the interband coupling on quantum transport, we have plotted in Fig. \ref{fig:probs_Z0_a05_rD15} the 
probabilities of the various reflection processes for an incoming electron from band $\lambda' = 1$ for the case of a transparent 
interface. For decoupled bands, all electrons are Andreev reflected into the same band (for subgap energies), and it is shown how 
this situation is altered for $\alpha > 0$ in a different manner for a $s_\pm$-wave superconductor and a two-band $s$-wave superconductor. 
The difference between the $s$-wave and $s_\pm$-wave case is reduced for increasing $Z$ relative to $\talpha$, and $|r_2|^2$ and 
$|r_2^A|^2$ are in general decreased by increasing $Z$ and increased by increasing $\talpha$. Interband scattering also effectively 
acts to reduce the interface transparency, although less so for the $s_\pm$ state. Apart from these general relations, the 
dependence of the probabilities on the coupling $\talpha$ is by no means trivial, and we do not attempt to give any further 
physical interpretation of this parameter.

The conductance for a two-band superconductor normalized to the normal state conductance $G_0$ may, within the BTK formalism, be given as 
\begin{equation}
	G/G_0 = \frac{1}{2F} \sum_{\lambda'} G_{\lambda'},
\end{equation}
with $G_\lambda' = 1 + |r_1^\text{A}|^2 + |r_2^\text{A}|^2 - |r_1|^2 - |r_2|^2$ for incoming electron in band $\lambda'$ (see Appendix A), and $F = 1 - |r_1|^2$ where the coefficient is evaluated for $|\Delta_1| = |\Delta_2| = 0$.

\begin{figure}[h]
	\centering
	\resizebox{0.45\textwidth}{!}{
	\includegraphics{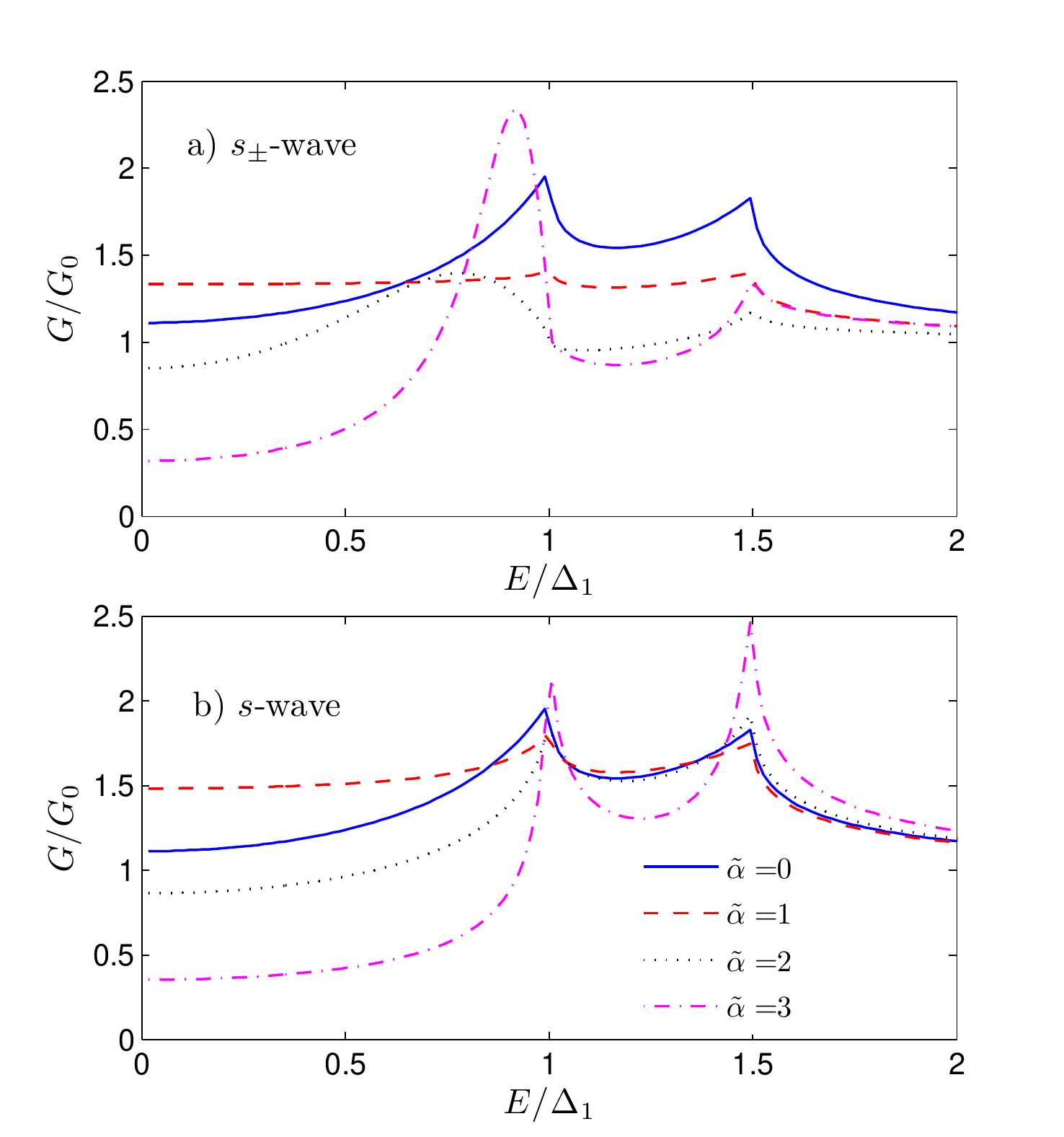}
	} \caption{(Color online) Conductance for a N$|s_\pm$-wave junction (a) and a two-band N$|s$-wave junction (b)  for various strengths of 
	interband coupling $\talpha$ normalized on its normal state value, where we have set $Z = 1$ and $r_\Delta = 1.5$.}
	\label{fig:cond_rD15}
\end{figure}

In panel (a) of Fig. \ref{fig:cond_rD15} we have plotted representative results for the conductance spectra for different values of the 
interband coupling. We have chosen the ratio between the gaps somewhat arbitrarily as $r_\Delta = |\Delta_2|/\Delta_1 = 1.5$, and have 
included the limiting case of $\talpha = 0$, which here simply corresponds to the well known BTK result with a double gap structure.
Furthermore, for values $\talpha > Z$ when $Z$ is small, the interband coupling enforces the formation of subgap peaks close to the 
gap edge which are damped and shifted to lower energies for decreasing $\talpha$. This feature becomes more prominent when 
$r_\Delta \rightarrow 1$ (not shown), which makes it observable also for larger $Z$, although also then in a restricted region 
of parameter space. As shown in panel (b) of Fig. \ref{fig:cond_rD15}, no features of this kind appear in the corresponding model 
without an internal phase shift in the superconductor.
In the conductance spectra of our model, we do not find the very strong low-energy conductance peaks reported 
in Ref. ~\onlinecite{golubov-tanaka_multiband_btk}, but rather features more reminiscent of those of 
Ref. ~\onlinecite{araujo-sacramento_multiband_andreev}, which may be reasonable since their approach was also based 
on the full BdG equations.

\subsection{Crossed Andreev Reflection}
\label{sec:car}

One of the most attractive prospects of CAR is as a realization of nonlocally correlated electron states, see 
\eg Ref. ~\onlinecite{falci-feinberg-hekking_car}. The CAR process is however often masked by the competing 
process of elastic cotunneling (EC), and it is therefore interesting to search for situations in which CAR 
dominates. In this section, we investigate how the internal phase difference of the $s_\pm$-wave state alters 
the nonlocal conductance.

For the left hand side lead ($x < 0$), we will use the same normal region wavefunction 
($\psi_{\rm N} \rightarrow  \psi_{\rm L}$) as in Eq. \eqref{eq:psiN}, and for the right hand 
side ($x> L$) lead we introduce
\begin{align}
	\label{eq:psiR}
	\psi_{\rm R} &= t_1 [1,0,0,0] \e{\i kx}  + t^A_1 [0,1,0,0] \e{-\i kx} \notag\\
	&+ t_2[0,0,1,0]\e{\i kx} + t^A_2[0,0,0,1] \e{-\i kx}.
\end{align}
For the superconducting interlayer ($0 < x < L$), we now have to rewrite the wave function of Eq. \ref{eq:psiS} 
into
\begin{align}
	\label{psiS2}
	\psi_S &= ( s_1 \e{\i q_1^+ x} + s_2 \e{-\i q_1^+ x} ) [u_1,v_1,0,0] + ( s_3 \e{\i q_1^- x} \notag\\
	&+ s_4 \e{-\i q_1^- x} ) [v_1,u_1,0,0] + ( p_1 \e{\i q_2^+ x} + p_2 \e{-\i q_2^+ x} ) \notag\\
	&[0,0,u_2,\delta v_2] + ( p_3 \e{\i q_2^- x} + p_4 \e{-\i q_2^- x} )  [0,0,\delta v_2,u_2],
\end{align}
where we have introduced the wavevectors
\begin{equation}
	q_\lambda^\pm = k_{\rm F} \sqrt{1 \pm \sqrt{E^2 - \Delta_\lambda^2} / E_{\rm F} },
\end{equation}
for electron- and hole-like quasiparticles, respectively. In the normal metal regions we can to a good approximation assume equal 
and constant wavevectors $k = k_{\rm F}$. In our calculations we have defined the Fermi energy by the value 
$E_{\rm F}/\Delta_1 = 10^4$.

We then apply the boundary conditions of Eq. \eqref{eq:bc} to the two interfaces at $x = 0$ and $x = L$, which results in $16$ 
equations in the variables $\{r_\lambda, r_\lambda^A, t_\lambda, t_\lambda^A, s_i, p_i\}$, which are solved numerically. 

\begin{figure}[h]
	\centering
	\resizebox{0.45\textwidth}{!}{
	\includegraphics{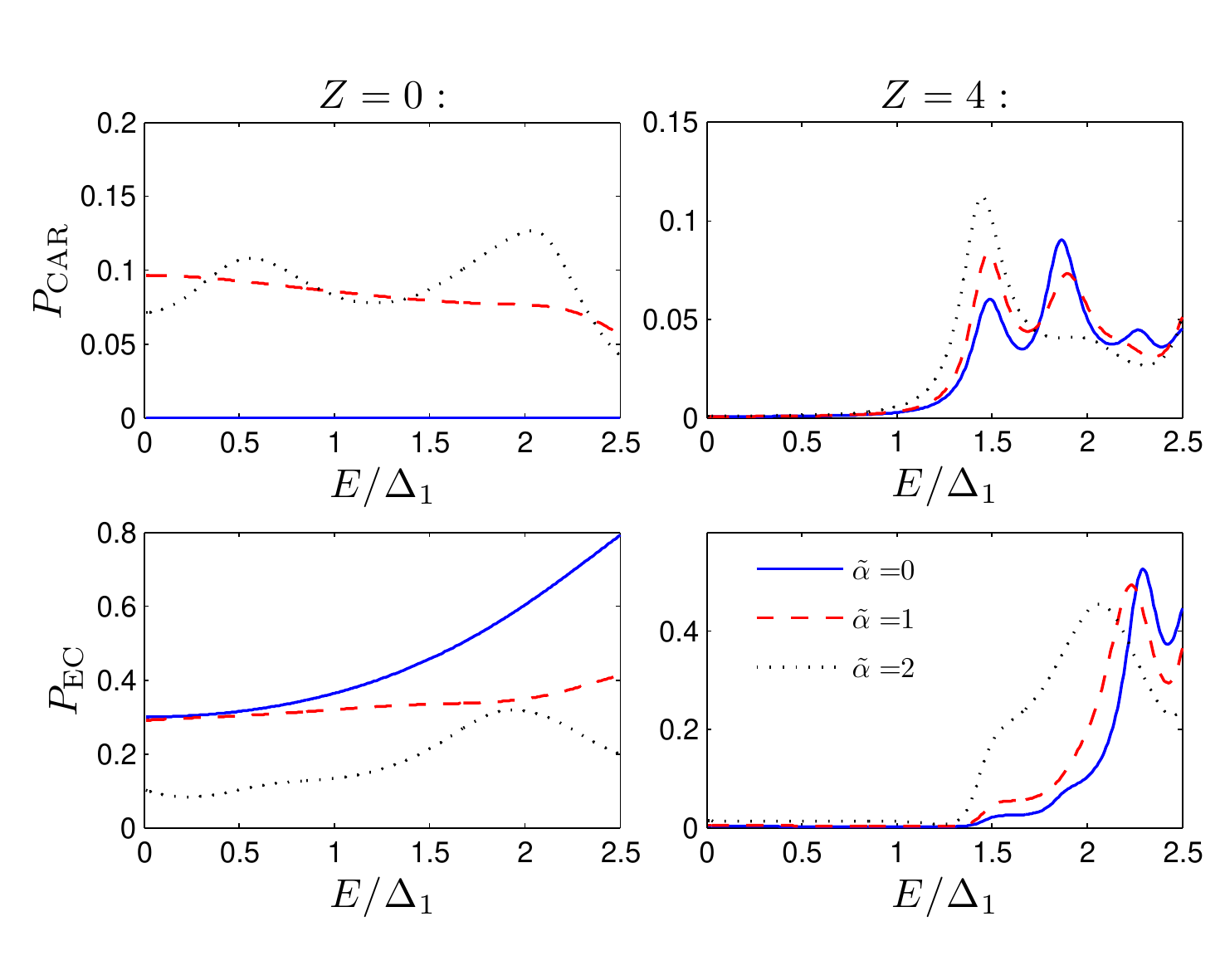}
	} \caption{(Color online) Nonlocal conductance through a N$|s_\pm$-wave$|$N junction for a relatively thin superconducting interlayer, 
	$L = 2 \cdot 10^4 k_{\rm F}^{-1}$, and for $r_\Delta = 1.5$. The upper panels show the probability measure for crossed Andreev 
	reflection, while the lower panels show for elastic cotunneling. We have used barrier strengths $Z = 0$ (left) and $Z = 4$ (right), 
	and a number of values for the interband coupling $\talpha$.}
	\label{fig:nonlocal_kL20K_rD15_Z0-4_3xa}
\end{figure}

Since it would have no physical meaning to measure the signal for the (virtual) normal metal bands $\lambda'$ separately, we 
choose to consider the average process probabilities
\begin{align}
	P_{\rm EC}  & = \frac{1}{2} \sum_{\lambda'} (|t_1|^2 + |t_2|^2), \\
	P_{\rm CAR} & = \frac{1}{2} \sum_{\lambda'} (|t^A_1|^2 + |t^A_2|^2), \\
\end{align} 
as the measure of nonlocal conductance, where $\sum_{\lambda'}$ again denotes summing over incoming electron bands. 

\begin{figure}[h]
	\centering
	\resizebox{0.45\textwidth}{!}{
	\includegraphics{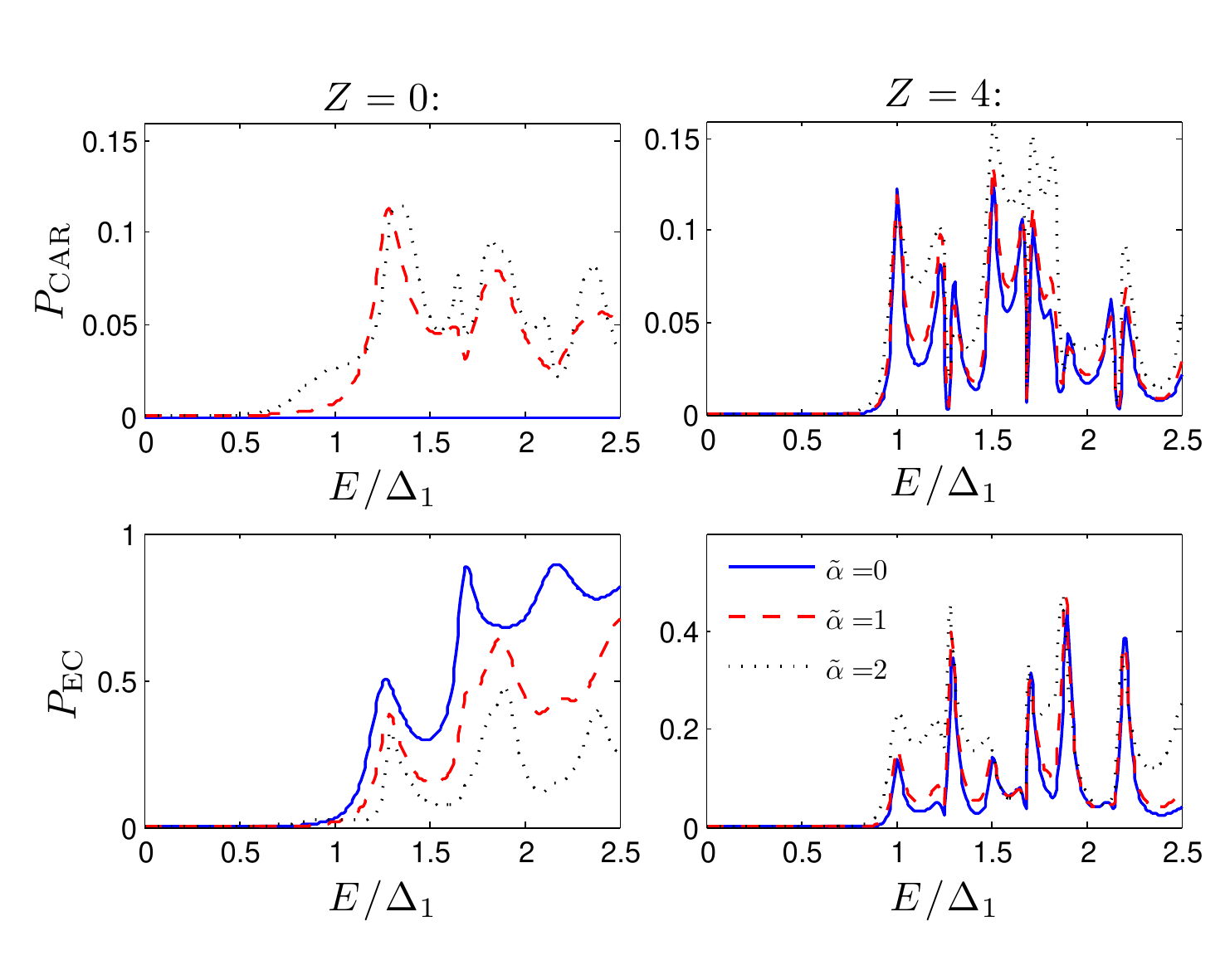}
	} \caption{(Color online) Nonlocal conductance through a N$|s_\pm$-wave$|$N junction for a relatively thick superconducting interlayer, 
	$L = 8 \cdot 10^4 k_{\rm F}^{-1}$, and for $r_\Delta = 1.5$. The upper panels show the probability  for crossed Andreev reflection, 
	while the lower panels show  the probability for elastic cotunneling. We have used barrier strengths $Z = 0$ (left) and $Z = 4$ 
	(right), and a number of values for the interband coupling $\talpha$.}
	\label{fig:nonlocal_kL80K_rD15_Z0-4_3xa}
\end{figure}

The nonlocal conductance is then proportional to $P_{\rm EC} - P_{\rm CAR}$, and we show the result for its separate 
contributions in Figs. \ref{fig:nonlocal_kL20K_rD15_Z0-4_3xa} and \ref{fig:nonlocal_kL80K_rD15_Z0-4_3xa}. As is 
expected for the components to the nonlocal conductance, it exhibits oscillations both as a function of energy and 
of the lead separation $L$, with decaying subgap contributions for increasing $L$.

It is seen that for high transparency, interband coupling facilitates the CAR process with respect to EC, a result which may be readily 
explained, since the coupling acts as an effective scattering barrier. Recall that for zero interface resistance (and no FWVM or spin polarization), the CAR process is completely absent. This result seems to be somewhat stronger 
for a $s_\pm$-wave superconductor than for a two-band $s$-wave superconductor (not shown), but $P_{\rm CAR}$ is never significantly 
larger than $P_{\rm EC}$. All in all, there are only minor qualitative differences to be found for the $s_\pm$-wave state when 
compared to a more conventional $s$-wave state, and we have therefore not included results for the latter here.

\subsection{Josephson current}
\label{sec:jos}

We now turn our attention to the Josephson coupling between two superconductors in a S$\mid$I$\mid$S junction with multiple bands. Below, 
we shall first consider the case where the right superconductor is $s_\pm$-wave while the left superconductor is single-band $s$-wave, 
with order parameter $\Delta_s = |\Delta_s| \exp{\phi_s}$. The strategy is to calculate analytically the Andreev bound 
states at the interface, which carry the Josephson current. These states are found by using the boundary conditions Eq. (\ref{eq:bc}) 
for the wavefunctions in each of the superconducting regions. However, since we will find that the interesting physics stems from allowing 
different band transmission, we let $V_0 \ \mathrm{diag}(\hat{1},\hat{1}) \rightarrow \hat{V} = \mathrm{diag}(V_1,V_1,V_2,V_2)$. For 
later reference, we also define $r_Z = Z_2/Z_1 = V_2/V_1$ as the ratio between the effective barrier strengths for the two bands; 
the motivation will be discussed in Sec. \ref{sec:discussion}. Using an alternative parameterization to that in Sec. \ref{sec:cond}, 
we write the wave function for the left hand side superconductor as
\begin{align}
	\psi_L &= s_1[1,\e{\i\beta_s},0,0] \e{-\i kx} + s_2[\e{\i\beta_s},1,0,0]\e{\i kx} \notag\\
	&+ s_3[0,0,1,\e{\i\beta_s}] \e{-\i kx} + s_4[0,0,\e{\i\beta_s},1]\e{\i kx},
\end{align}
while we for the right superconducting region have
\begin{align}
	\psi_R &= t_1[1,\e{\i(\beta_1 - \varphi)},0,0] \e{\i kx} + t_2[\e{\i(\beta_1 + \varphi)},1,0,0]\e{-\i kx} \notag\\
	&+ t_3[0,0,1,\delta \e{\i(\beta_2 - \varphi)}] \e{\i kx} + t_4[0,0,\delta \e{\i(\beta_2 + \varphi)},1]\e{-\i kx},
\end{align}
where $\beta_s = \arccos{(E/|\Delta_s|)}$ and $\beta_\lambda = \arccos{(E/|\Delta_\lambda|)}$. The gauge invariant phase 
difference between the two superconductors has been defined as $\varphi = \phi_1 - \phi_s$. 

Setting up the boundary conditions of Eq. \eqref{eq:bc} yields a system of equations on the form
\begin{equation}
\label{eq:det_is_zero}
	\Lambda \mathbf{t} = \mathbf{0},
\end{equation}
where $\mathbf{t} = \{s_1^\text{L}, t_1^\text{L}, s_2^\text{L}, t_2^\text{L}, s_1^\text{R}, t_1^\text{R}, s_2^\text{R}, t_2^\text{R} \}$ 
and $\Lambda$ is a $8 \times 8$ matrix. The Andreev bound states are found by requiring a nontrivial solution for the system,
$\text{det}(\Lambda) = 0$, which in general results in four energy states $E_\lambda^\pm(\varphi)$. The Josephson current is found 
in the ordinary way by\cite{golubov_cpr_rmp}
\begin{equation}
\label{eq:current}
	I = 2 e \sum_{i=1}^4 \pdiff{E_i}{\varphi} f(E_i),
\end{equation}
where $E_i$ denotes the four ABS and $f(E)$ is the Fermi-Dirac distribution function. We will define the critical current $I_c$ as 
the maximal current allowed by the current-phase relation for a given set of parameters, $I_c = \text{max}\{I(\varphi)\}$. We also 
introduce the quantity $I_0 = 2 e |\Delta_1|$ used for normalization of the current.

\subsubsection{0-$\pi$ phase shifts for varying barrier strengths: the case of equal gap magnitudes}
\label{sec:equalgaps}

\begin{figure}[h]
	\centering
	\resizebox{0.45\textwidth}{!}{
	\includegraphics{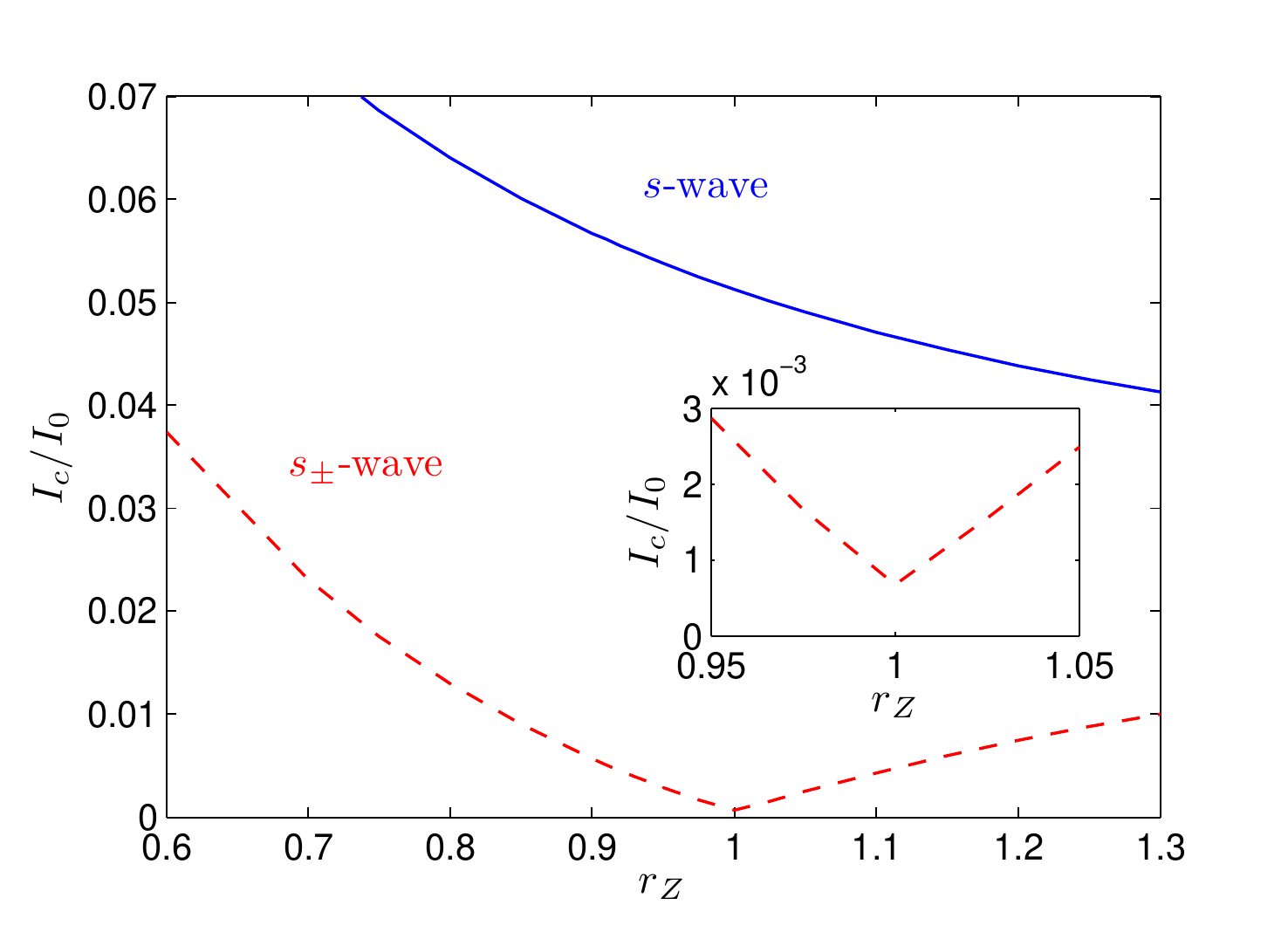}
	} \caption{(Color online) Critical current for a $s$-wave$|$I$|$$s_\pm$-wave Josephson junction as a function of the ratio $r_Z$ 
	between the effective barrier strength of band 2 and band 1. There is no interband coupling ($\talpha = 0$), and we have 
	set $Z_1 = 6$ and $T = 0$. Also shown is the critical current in the case that the right hand side is a (two-band) $s$-wave 
	superconductor.}
	\label{fig:Ic_vs_rZ}
\end{figure}

\begin{figure}[h]
	\centering
	\resizebox{0.45\textwidth}{!}{
	\includegraphics{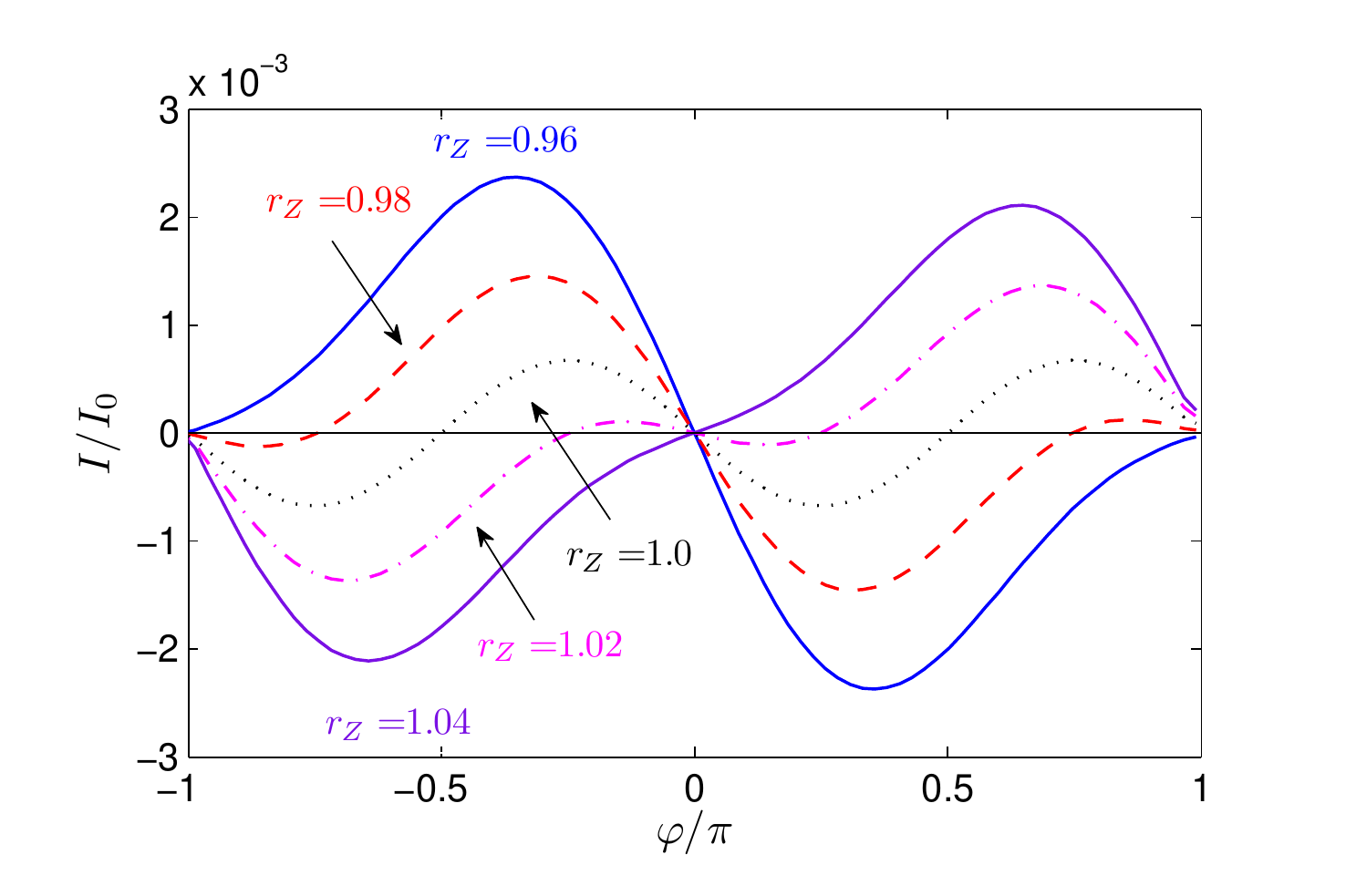}
	} \caption{(Color online) Current-phase relation for a $s$-wave$|$I$|$$s_\pm$-wave Josephson junction with zero 
	interband coupling ($\talpha = 0$), with $Z = 6$ and $T = 0$.}
	\label{fig:CPR_rS1_rD1_Z6_5xrZ}
\end{figure}

Before investigating the temperature dependence of the critical current, we will consider the limit $T \rightarrow 0$. Our main results 
in this section is the observation of a 0-$\pi$ transition in the Josephson current for varying the barrier strength ratio $r_Z$, as 
shown in Fig. \ref{fig:Ic_vs_rZ}. For a $s$-wave$|$I$|$$s_\pm$-wave Josephson junction this can be understood in a very simple manner 
as the competition between the $\lambda = 1$ and the $\lambda = 2$ band components of the current; the band with order parameter 
$\Delta_1 =|\Delta| \e{\i\phi_1}$ will favor the conventional 0-junction whereas at the same time the other band with 
$\Delta_2 = -|\Delta| \e{\i\phi_1}$ will favor a $\pi$-junction. Here, we have for simplicity assumed that 
$|\Delta_\lambda| = |\Delta_s| \equiv |\Delta|$. To show this mechanism explicitly we proceed analytically in the limit of 
$\talpha = 0$, and this minimal model also serves as a review of the basic physics involved in a ballistic Josephson junction. 
Now, the solutions for Eq. \eqref{eq:det_is_zero} can be shown to be
\begin{align}
\label{eq:ABS}
	E_1^\pm = \pm |\Delta| \sqrt{1 - D_1 \sin^2(\varphi/2)},\\
	E_2^\pm = \pm |\Delta| \sqrt{1 - D_2 \cos^2(\varphi/2)},\notag
\end{align}
where $D_\lambda = 4 / (4 + Z_\lambda^2)$. $E_1^\pm$ are the well known solutions for a one-band $s$-wave$|$I$|s$-wave 
junction\cite{golubov_cpr_rmp}, while $E_2^\pm$ are the corresponding solutions for the negative-gap band. Expanding to 
first order in $D_\lambda$ and inserting in Eq. \eqref{eq:current} yields the Josephson current
\begin{equation}
	I = I_1 \sin{\varphi},
\end{equation}
where $I_1 =(D_1 - D_2) I_0 / 4$. It is obvious that for $Z_2 < Z_1$ one will have $D_2 > D_1$ and $I_1 < 0$, \ie the system 
being in the $\pi$ state. As shown in Fig. \ref{fig:Ic_vs_rZ}, the crossover point above which the $\lambda = 1$ contribution 
dominates instead is $r_Z = 1$. However, inspection shows that the current does not vanish entirely at the crossover point, a 
fact which is readily explained by going to the second order expansion of Eq. \eqref{eq:ABS}. In the limit $Z_2 \rightarrow Z_1$ 
partial cancellation of the two 1st order terms then reduces the current to
\begin{equation}
	I = I_2 \sin(2\varphi),
\end{equation}
where $I_2 = - I_0 D_\lambda^2/ 16$. In other words, the second harmonic component to the current appears, and is dominating 
close to the transition point. The general non-sinusoidality of the current-phase relation close to the transition point is 
illustrated in Fig. \ref{fig:CPR_rS1_rD1_Z6_5xrZ}.

Before proceeding, it will be instructive for the subsequent discussion to analyze this current-phase relation a little 
further. In a region close to $r_Z  = 1$  (and for relatively large Z), we may write out the approximate Josephson current 
to be given by the expression
\begin{equation}
	\label{CPR_approx}
	I/I_0 = \frac{D_1 - D_2}{4} \sin{\varphi} - \frac{(D_1 + D_2)^2}{64} \sin{(2 \varphi)}.
\end{equation}
For a Josephson junction containing a second harmonic component in the current-phase relation, the ground state needs neither 
to be a 0-state nor a $\pi$-state, but may instead be a $\varphi$-state
\cite{buzdin-koshelev_alternating_0-pi_junction, buzdin_0_to_pi-junction} with a general equilibrium phase difference 
$\varphi_0$. This ground state phase can for our case be found as\cite{goldobin_2nd_harmonic}
\begin{equation}
	\varphi_0 = \arccos{\left( \frac{8(D_1-D_2)}{(D_1+D_2)^2} \right)}.
\end{equation}
This phase value evolves smoothly from $\varphi_0 = \pi$ for $r_Z \ll 1$ to $\varphi_0 = 0$ for $r_Z \gg 1$, passing 
$\varphi_0 = \pi/2$ at $r_Z = 1$. For the case of $Z = 6$, our model system is a $\varphi$-junction for an approximate 
region $r_Z \in (0.97, 1.028)$, and we have verified numerically that Eq. \eqref{CPR_approx} is qualitatively a very 
good approximation also well outside this region. The phase difference which supports the critical current will on the 
other hand be denoted as $\varphi^\ast$, and can in a similar manner be found to evolve from to $-\pi/2$ for the $\pi$-state 
at $r_Z \ll 1$ to $\varphi^\ast = -\pi/4$ for $r_Z = 1^-$, where it jumps discontinuously to $\varphi^\ast = 3\pi/4$ for 
$r_Z = 1^+$, from which it again evolves smoothly towards $\pi/2$ for the limiting sinusoidal current-phase relation. 
This phase-shift mechanism will be instrumental to the findings presented in Sec. \ref{sec:diffgaps}.
 
\begin{figure}[h]
	\centering
	\resizebox{0.45\textwidth}{!}{
	\includegraphics{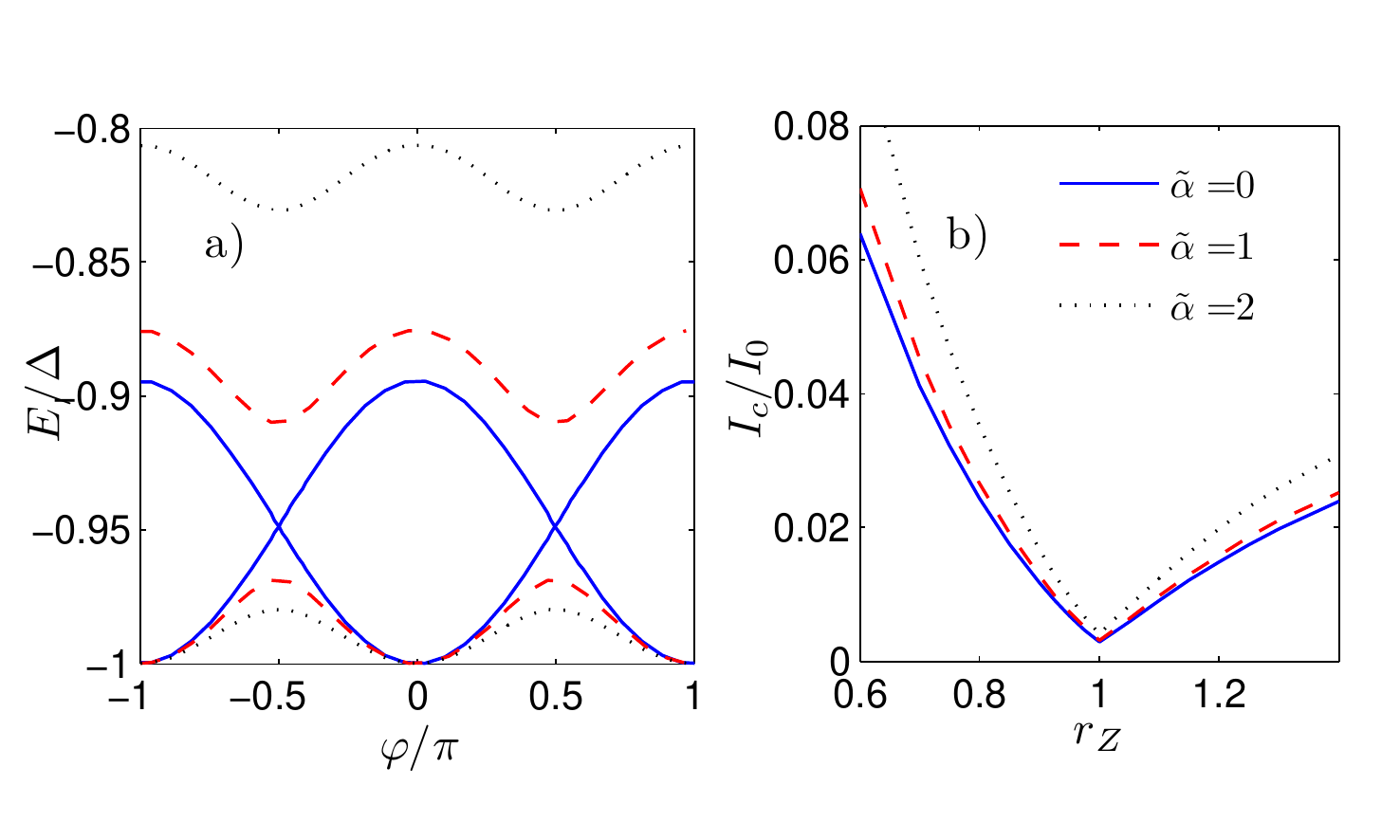}
	} \caption{(Color online) Josephson coupling for a $s$-wave$|$I$|$$s_\pm$-wave Josephson junction for various values of 
	interband coupling strength $\tilde{\alpha}$, with a) dispersion of the two Andreev bound states with $E < 0$ shown to 
	the left and b) critical current as a function of barrier strength ratio $r_Z$ shown to the right. Both results are 
	given for intermediate barrier strength $Z = 4$.}
	\label{fig:ABS_and_Ic_vs_rZ_Z4_3xa}
\end{figure}

Next we investigate the general case with nonzero interband coupling. The (numerical) solution for the two lower ABS energies at 
zero temperature is shown in panel (a) of Fig. \ref{fig:ABS_and_Ic_vs_rZ_Z4_3xa} for different values of $\talpha$. While the 
energy states cross each other according to Eq. \eqref{eq:ABS} for $\talpha = 0$, they repel each other for nonzero interband 
coupling, forming a gap in the ABS energy spectrum which increases for increasing $\talpha$. As a trivial observation, this 
can be understood as a hybridization of the two formerly independent bands, as a finite hopping term introduces off-diagonal 
matrix elements in $\lambda$-space. The general properties of the current-phase relation remains the same in spite of the 
explicit $\pi$-periodicity of the ABS dispersion, and also here this is explained by the partial cancellation of the two 
ABS contributions. As can be seen from panel (b) of Fig. \ref{fig:ABS_and_Ic_vs_rZ_Z4_3xa}, $\talpha > 0$ does not change 
the behavior of the Josephson current in any dramatic way, and neither does the interband coupling influence the position 
of the 0-$\pi$ transition point; it remains at $r_Z = 1$ for all values of $\talpha$. This motivates us to suppress the 
interband coupling $\talpha$ in what follows to be able to obtain analytically tractable results.

\subsubsection{Magnetic field dependence of the critical current}
\label{sec:fraunhofer}

As a simple application of the model described in the preceding section, we now calculate the dependence of the critical current 
$I_c$ on an external magnetic field $H$, \ie the magnetic diffraction pattern. This quantity is experimentally very interesting, 
and experimental results for $I_c(H)$ have recently been presented for iron-based superconductors
\cite{zhou_iron_fraunhofer, zhang_iron_josephson, zhang_iron_josephson2}. For our model, we are interested in studying how the magnetic diffraction 
patterns depend on the relative barrier strength of the two bands, as $r_Z$ is seen as the primary parameter determining the 
behavior of the system.

\begin{figure}[h]
	\centering
	\resizebox{0.45\textwidth}{!}{
	\includegraphics{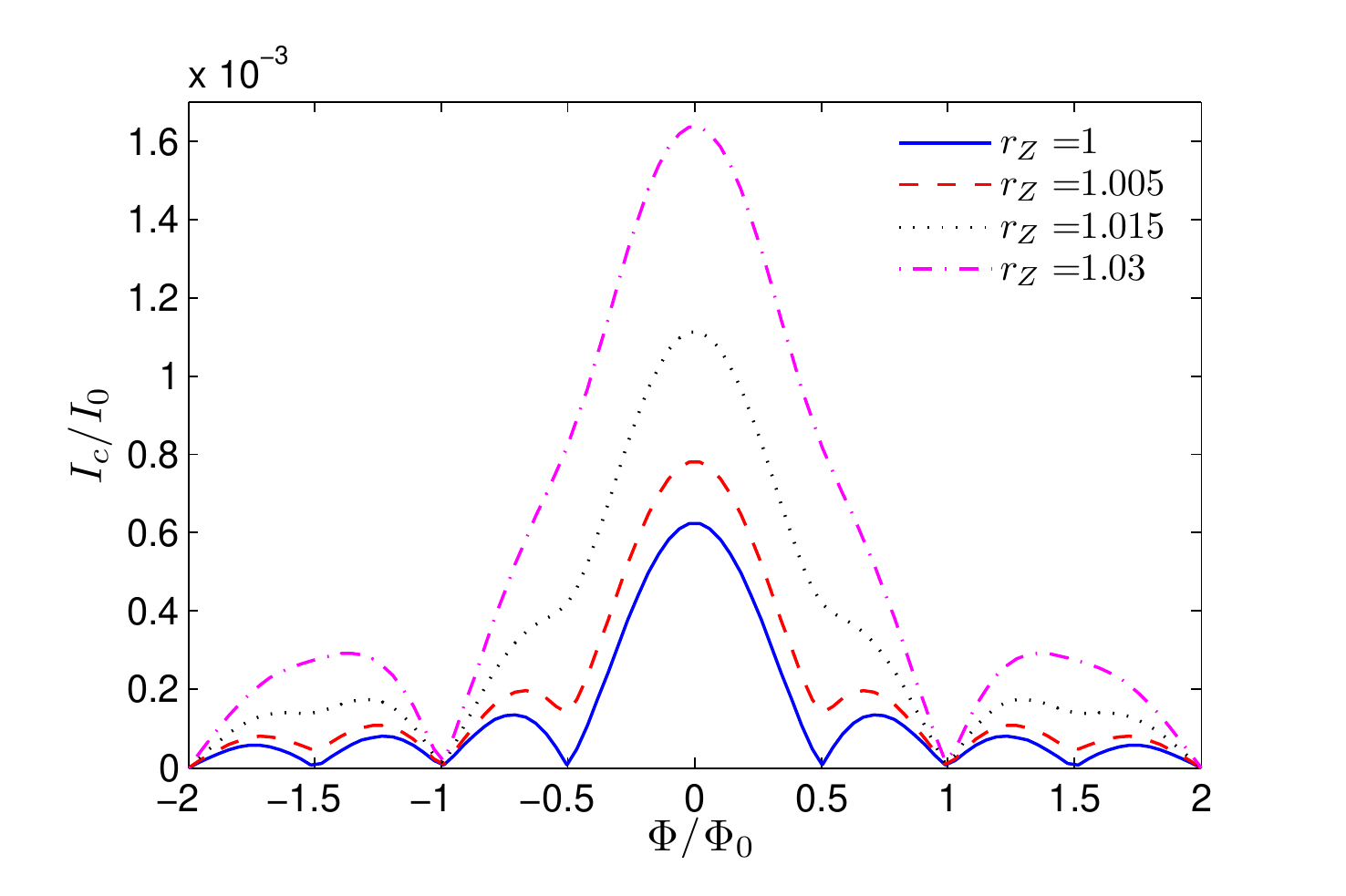}
	} \caption{(Color online) Fraunhofer-like magnetic diffraction pattern for a $s$-wave$|$I$|$$s_\pm$-wave junction in an external magnetic field for various values of barrier strength ratio. We have used the parameter values $Z=6$, $\talpha=0$.}
	\label{fig:fraunhofer_Z6_4xrZ}
\end{figure}

In order to include an external magnetic field to our model system, we must define a width $W$ along the $z$-axis and an effective 
length $d_{\rm J}$ around $x=0$ over which the magnetic field $H$ along the $y$-axis penetrates the junction. The magnetic flux 
through the junction is then given by $\Phi = H W d_{\rm J}$, and we let $\Phi_0$ denote the magnetic flux quantum. Using 
the approximation of Eq. \eqref{CPR_approx} for the current-phase relation, we can study our system in the framework of 
Ref. ~\onlinecite{goldobin_2nd_harmonic}, from which we straightforwardly find the expression
\begin{align}
	I_c(\Phi) = \Big[ \frac{D_1 - D_2}{4} \sin{ \Big(\frac{\pi \Phi}{\Phi_0} \Big) }  \sin{\varphi} \notag\\
	- \frac{(D_1 + D_2)^2}{128} \sin{ \Big( \frac{2\pi \Phi}{\Phi_0} \Big) } \sin{(2\varphi)} \Big]/\Big[ \frac{\pi \Phi}{\Phi_0} \Big].
\end{align}
Evaluating the above expression for the phase difference $\varphi = \varphi^\ast$ giving the maximum current for the respective $r_Z$, 
we obtain the Fraunhofer-like diffraction patterns shown in Fig. \ref{fig:fraunhofer_Z6_4xrZ}. The effect of the 2nd harmonic 
component to the current is evident as a half-integer flux quantum modulation of the critical current which grows more pronounced 
as $r_Z \rightarrow 1$, but whose contribution is vanishing outside the $\varphi$-junction region. Although results for $r_Z < 1$ 
are not shown here, these are largely symmetric with respect to $r_Z = 1$. We may also note that similar results for the magnetic 
diffraction were presented Ref. ~\onlinecite{buzdin-koshelev_alternating_0-pi_junction}, albeit for a completely different 
system.

\subsubsection{Temperature dependence of the Josephson current: the case of different gap magnitudes}
\label{sec:diffgaps}

Motivated by the indications in Ref. ~\onlinecite{linder_iron_rapid} that different gap magnitudes are necessary for the occurrence of thermally induced 0-$\pi$-transitions, we now consider a system for the general case of $\Delta_s \neq \Delta_1 \neq |\Delta_2|$. As we showed in Sec. \ref{sec:equalgaps}, interband coupling did not affect the 0-$\pi$-transitions as a function of $r_Z$ qualitatively, so we will assume in the following that $\alpha = 0$, an approximation which moreover makes an analytical approach feasible. Solving the $8\times8$ system as two decoupled $4\times4$ systems, we obtain the analytical solution as given by Eqs. \eqref{eq:diffgaps1} and \eqref{eq:diffgaps2} in Appendix B. We also refer to this appendix for some more information regarding validity, existence and uniqueness of this solution. 

\begin{figure}[h]
	\centering
	\resizebox{0.45\textwidth}{!}{
	\includegraphics{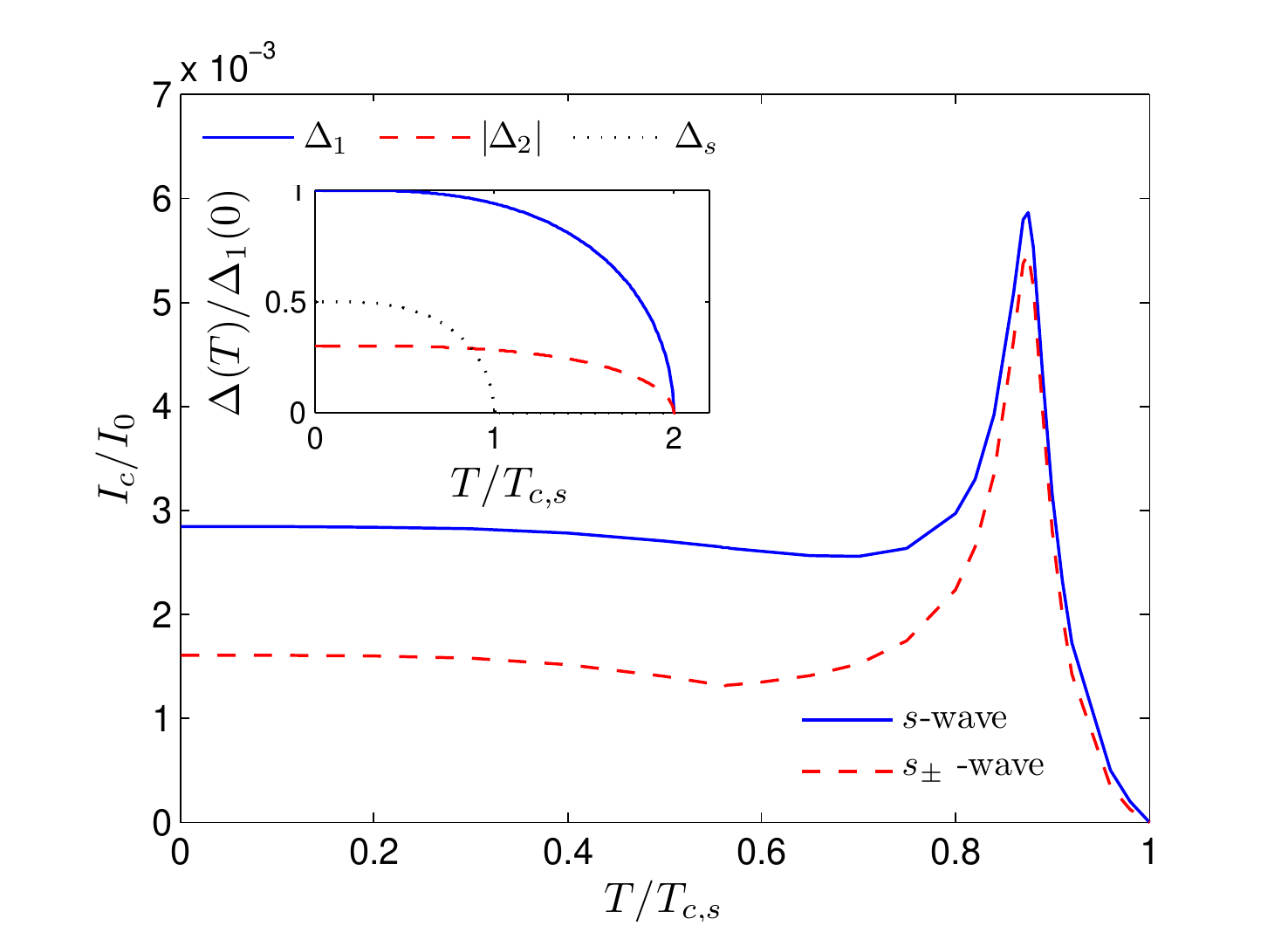}
	} \caption{(Color online) Temperature dependence of the critical current for the parameters $r_s = 0.5$, $r_\Delta = 0.3$, with $Z = 6$ and $r_Z = 1$. Both the results for a two-gap $s$-wave and a $s_\pm$-wave superconductor are shown. \textit{Inset:} temperature dependence of gap magnitudes for the same parameter set.}
	\label{fig:Ic_vs_T_rS05_rD03_2xd_inset}
\end{figure}

We will assume BCS-like temperature dependence of the gaps, with the $s$-wave gap of the left superconductor closing at a temperature $T_{c,s} = \Delta_s(T=0)/1.76$ while both gaps of two-band superconductor on the right hand sides close simultaneously at $T_{c,\lambda} \equiv T_c = \Delta_1(T=0)/1.76$ . We will parameterize the difference in gap magnitudes by $r_s = \Delta_s/\Delta_1$ and $r_\Delta = |\Delta_2|/\Delta_1$, and will in most of what follows restrict ourselves to $r_s = 0.5$ and $r_\Delta = 0.3$ as a representative set of gap ratios, although we stress that our results are valid in a much larger portion of parameter space. The resulting temperature dependence of the three superconducting gaps is illustrated in the inset of Fig. \ref{fig:Ic_vs_T_rS05_rD03_2xd_inset}.

\begin{figure}[h]
	\centering
	\resizebox{0.45\textwidth}{!}{
	\includegraphics{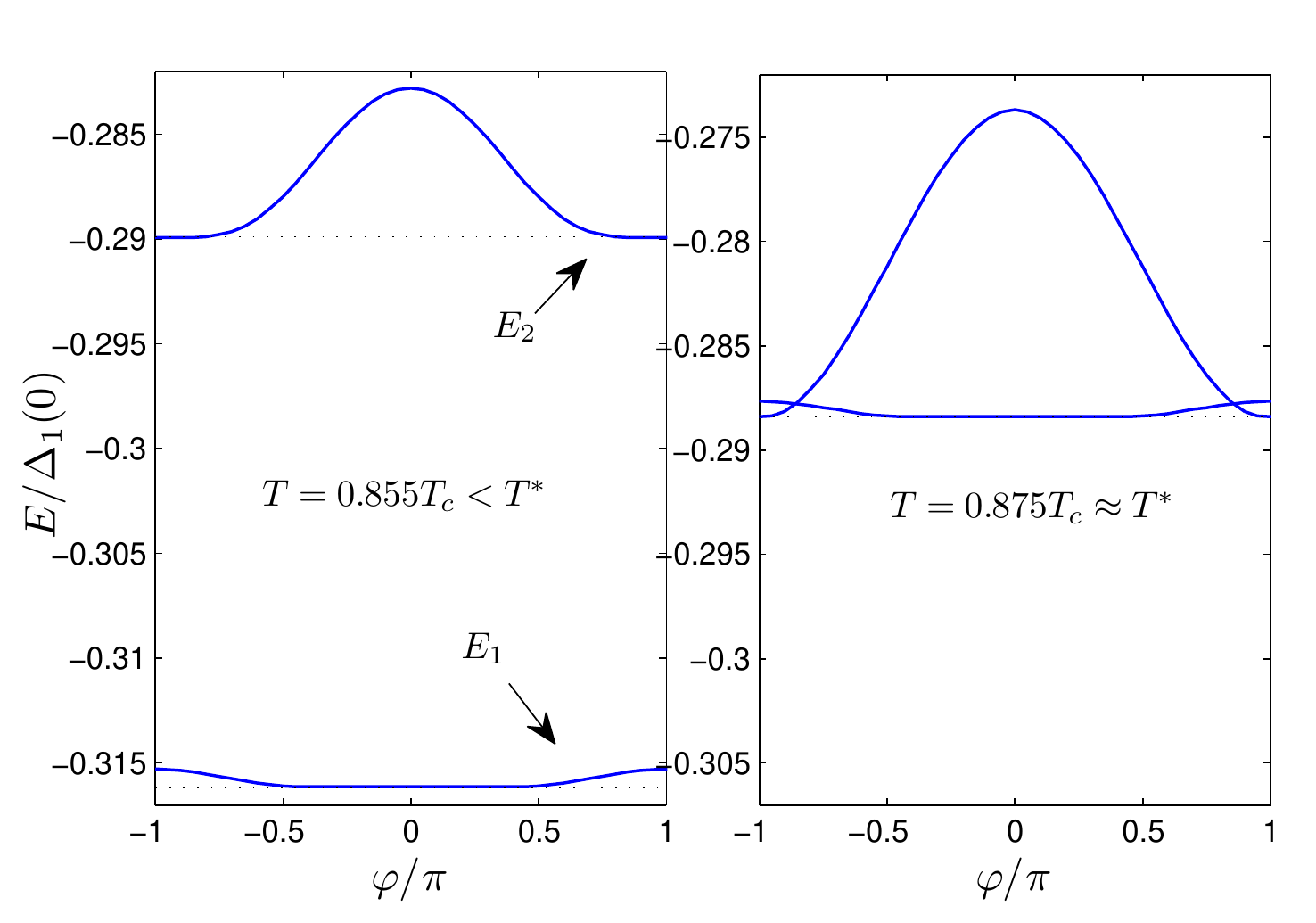}
	} \caption{(Color online) Energy of Andreev bound states $E_\lambda^-$ for two temperatures close to the point where $|\Delta_2(T)| = \Delta_s(T)$ for $r_s = 0.5$ and $r_\Delta = 0.3$. Other parameters are $Z = 6$ and $r_Z = 1$. Shown with dotted lines are the relevant gap edges which the energy states track.}
	\label{fig:ABS_rS05_rD03_T0855-0875}
\end{figure}

First, we compare the temperature dependence of the critical current both for a $s_\pm$-wave and a two-gap $s$-wave superconductor in Fig. \ref{fig:Ic_vs_T_rS05_rD03_2xd_inset}. The most distinctive feature for both these cases is the sharp peak at high temperature. This is exactly the temperature $T = T^\ast$ for which two of the gaps cross, \ie $|\Delta_2(T^\ast)| = \Delta_s(T^\ast)$, and although this peak is not a signature of the $s_\pm$-state as such, since it is present irrespective of the phase difference between the two right hand side gaps, it would be interesting to disclose the mechanism behind this feature. We turn therefore to the energy dispersion of the ABSs, as shown in Fig. \ref{fig:ABS_rS05_rD03_T0855-0875}  for two temperatures close to the peak in the critical current. Firstly, this illustrate how $E_2$ tracks the gap edge of $|\Delta_2(T)|$ and $E_1$ the gap edge of $\Delta_s(T) < \Delta_1(T)$ as the temperature is increased, whereas for $T > T^\ast$ both states track the smallest of the gaps, \ie $\Delta_s(T)$. Secondly, we observe that the energy states are non-dispersive for a phase interval centered around $\varphi = 0$ and $\varphi = \pm \pi$, for $E_1$ and $E_2$, respectively (\cf the discussion in Appendix B), so that in these regions the current contributions of the states vanish. Thirdly, we also observe that the dispersion of $E_2$ is strongly enhanced at $T = T^\ast$. This last observation can be understood by glancing at the expression for $E_2$ in Eq. \eqref{eq:diffgaps2} for a given $T$, from which we realize \eg by setting $\cos{(\varphi/2)} = 1$, that the band width of the energy state is at its maximum for $\Delta_s = |\Delta_\lambda|$. This is of course exactly the case for $T = T^\ast$. Moreover, since the contribution from $E_1$ vanishes for a large $\varphi$-interval for this temperature, whereas it is non-vanishing for $E_2$ for all $\varphi$ in this limiting case of equal gap magnitudes, one does not get the effect of partial cancellation of the two current contributions that was present for lower temperatures and for $\Delta_s = \Delta_1 = |\Delta_2|$. We note that although the peak strength for these gap ratios is somewhat extreme, we have verified that similar peaks or bumps persists in a major part of parameter space. 

\begin{figure}[h]
	\centering
	\resizebox{0.45\textwidth}{!}{
	\includegraphics{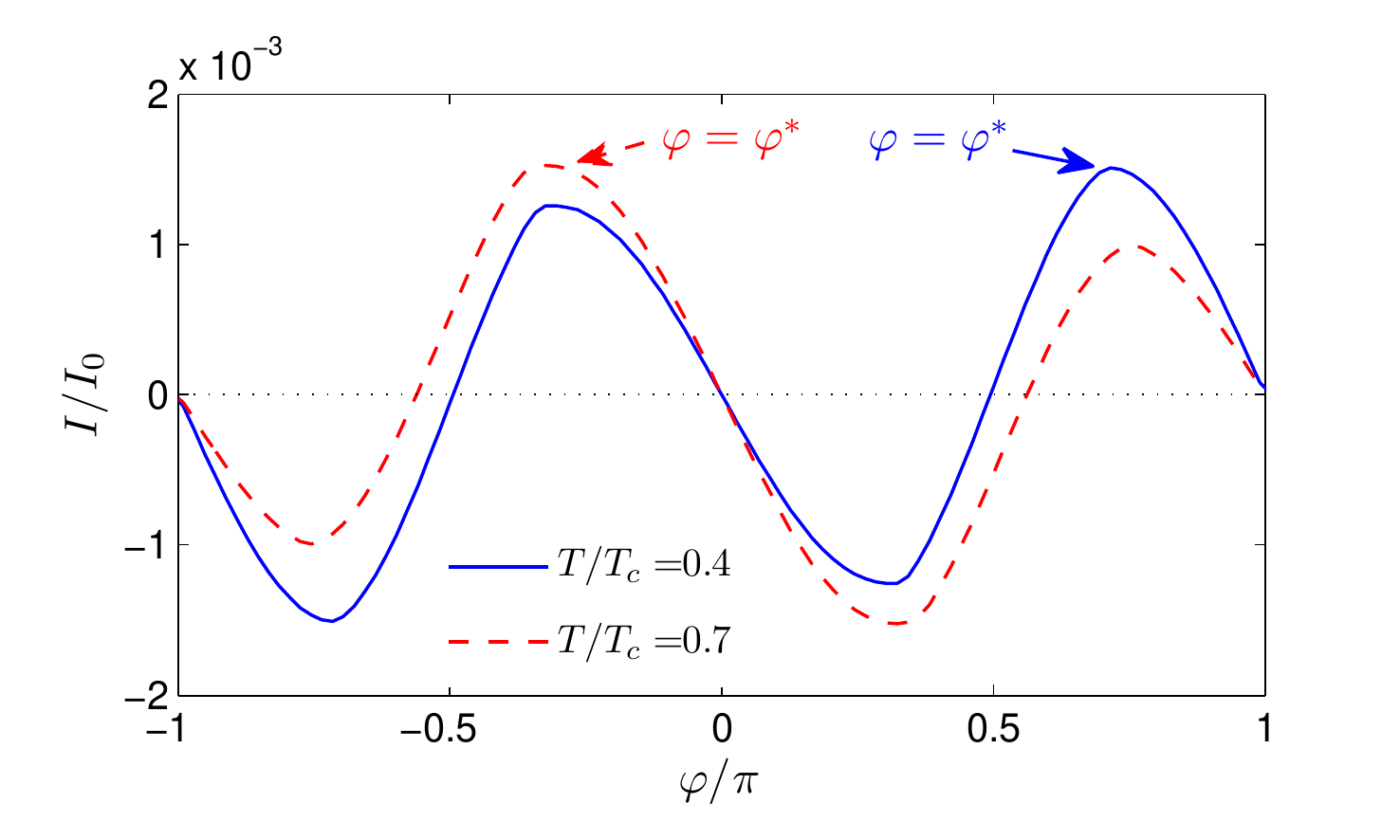}
	} \caption{(Color online) Current-phase relation for a system slightly below ($T/T_c = 0.4$) and slightly above ($T/T_c = 0.7$) the thermally induced phase shift appearing in the Josephson junction with gap ratios $r_s = 0.5$, $r_\Delta = 0.3$ and with $Z = 6$ and $r_Z = 1$. The arrows indicate the phase difference supporting the critical current ($I>0$) for the two temperatures.}
	\label{fig:CPR_rS03_rD05_T04-07}
\end{figure}

We have concluded that a peak in the critical current cannot be taken as a signature of $s_\pm$-pairing since it results from the energy gap crossing of the right hand and left hand superconductor in general. We therefore return to our investigation into possible thermally induced 0-$\pi$ phase shifts as an unambiguous sign of $s_\pm$-wave pairing, although apparently no such phase shift is present in our results. However, we remember from the analysis of the current-phase relation in Sec. \ref{sec:equalgaps} that in the presence of a second harmonic component to the current, a prospective 0-$\pi$ transition was smeared out into a $\varphi$-state region for which the critical current remains nonzero. We therefore consider the current-phase relation for the junction with different gap magnitudes in Fig. \ref{fig:CPR_rS03_rD05_T04-07} for two intermediate temperature values. It is evident that the second-harmonic component dominates, a fact which can be traced back to the vanishing of the ABS contributions for complementary phase intervals as discussed above. A related result is that the two maxima shown in Fig. \ref{fig:CPR_rS03_rD05_T04-07} originate predominantly from one of the energy states each. Furthermore, we have seen that the contribution from an ABS is larger the closer the values of the gap magnitudes $\Delta_s(T)$ and $|\Delta_\lambda(T)|$, and as $T \rightarrow T^\ast$, $\Delta_s(T)$ and $|\Delta_2(T)|$ are closing in on each other whereas $\Delta_s(T)$ and $\Delta_1(T)$ are moving apart. Thus the difference in the rate at which the gaps decrease causes the $E_1$ state to lose dominance to $E_2$ for increasing temperature. (Since $\Delta_1$ is by far the largest of the gaps, the corresponding ABS dominates for $T=0$ even though $\Delta_1$ is further from $\Delta_s$ than is $|\Delta_2|$.)

\begin{figure}[h]
	\centering
	\resizebox{0.45\textwidth}{!}{
	\includegraphics{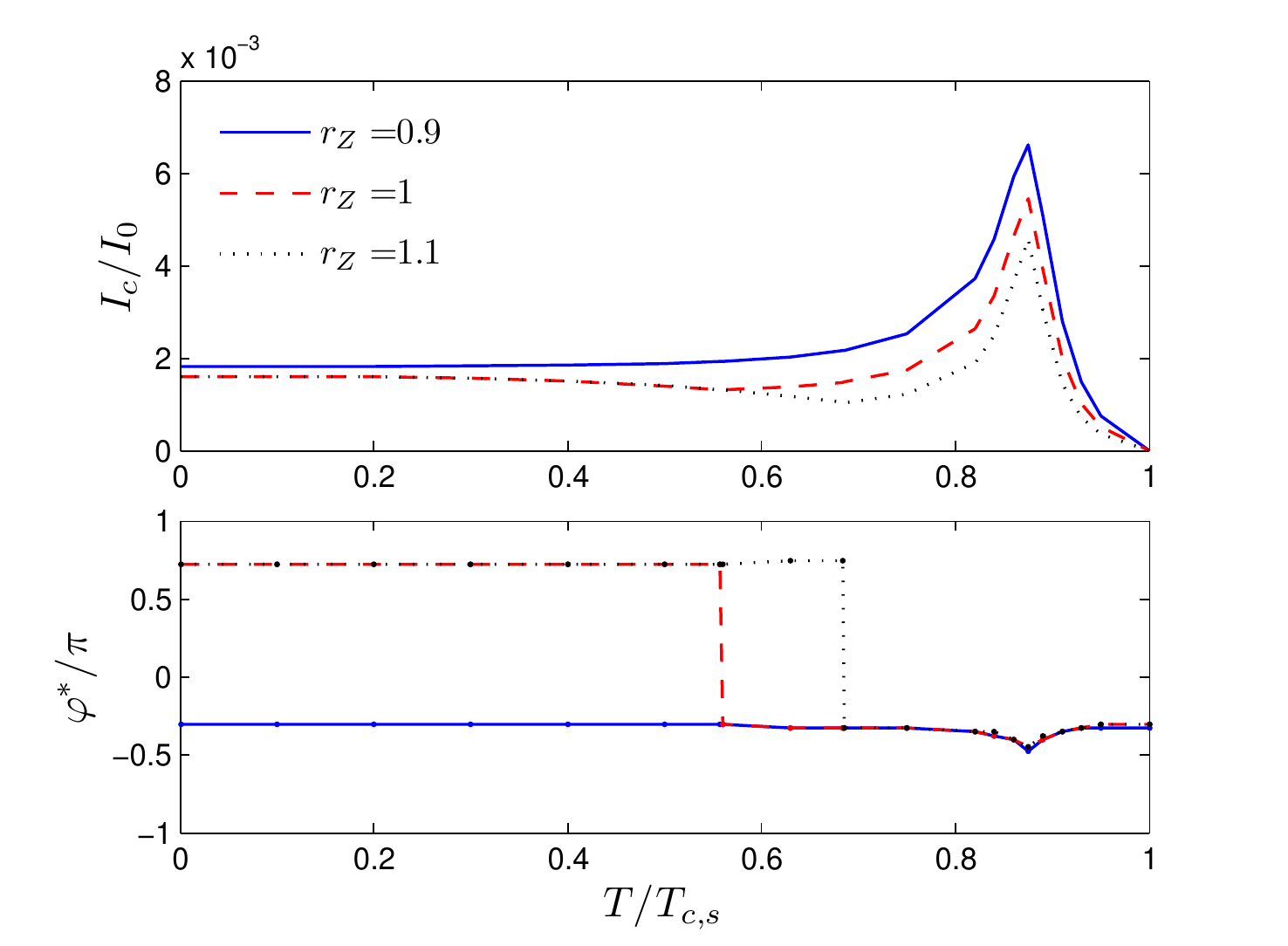}
	} \caption{(Color online) The upper panel shows the temperature dependence of the critical current for the parameters $r_s = 0.5$, $r_\Delta = 0.3$ and $Z=6$, similarly as in Fig. \ref{fig:Ic_vs_T_rS05_rD03_2xd_inset}, but for various values of barrier strength ratios $r_Z$. The lower panel shows the phase difference $\varphi^\ast$ supporting the critical current as a function of temperature for the same parameter set as above, illustrating the effect of the (discontinuous) phase shift from $\varphi^\ast > 0$ to $\varphi^\ast < 0$.}
	\label{fig:Ic_vs_T_rS05_rD03_w_phi}
\end{figure}

In Fig. \ref{fig:CPR_rS03_rD05_T04-07} we have also indicated the phase difference $\varphi^\ast$ in the current-phase relation that supports the critical current for each of the two temperatures. We now understand that as the dominant contribution to the current changes from $E_1$ to $E_2$ with increasing temperature, there must be a jump in this phase value from $\varphi^\ast > 0$ to $\varphi^\ast < 0$, and this jump needs to happen discontinuously at the temperature $T = T_\varphi$ where the two contributions balance (\cf our discussion of $I_c(r_Z)$ in Sec. \ref{sec:equalgaps}). This is our main result in this section: Although the Josephson junction is at no point in a 0-state or a $\pi$-state, the system may nevertheless exhibit discernible phase shifts when residing in the $\varphi$-state. We illustrate this phenomenon for different parameters in Fig. \ref{fig:Ic_vs_T_rS05_rD03_w_phi}, and note that similar behavior was observed for a large set of different gap ratios as long as $r_s \neq 1$ and $r_\Delta \neq 1$, the basic mechanism behind it being different temperature dependence of the different gaps. For the case of a two-gap $s$-wave state, a phase shift is of course not possible, as the two contributions to the current are then acting cooperatively at all times.

\section{Discussion}
\label{sec:discussion}

Comparing the three systems considered in the previous section, it is easy to see that role played by interband scattering differs fundamentally. 
On the one hand, tunneling spectroscopy and nonlocal conductance in the absence of interband coupling is not dependent upon the relative 
phase difference of the two $s_\pm$-wave order parameters, being merely the sum the contribution from two decoupled $s$-wave states. On 
the other hand, phase information enters explicitly into the calculation of the Josephson current, so that the interplay between the 
phases of the two order parameters is evident also for zero interband coupling. Furthermore, it seems that the behavior observed for 
the Josephson current remains qualitatively unaltered also for finite $\alpha$. This explains how it seems much more appealing to 
obtain phase information from multiband superconductors by the use of Josephson junctions than by tunneling spectroscopy, and why 
we will focus our discussion on this experimental probe.

To be able to compare our results for the ballistic limit with our previously obtained results for the diffusive limit in 
Ref. ~\onlinecite{linder_iron_rapid}, we now briefly recapitulate this work. Here we employed the quasiclassical Usadel 
equation\cite{usadel} to study Josephson coupling in a $s$-wave$|$N$|s_\pm$-wave junction in the limit of weak proximity 
effect, an approximation which is warranted for low-transparency interfaces. We showed that for this case, 0-$\pi$ transitions 
were observed both as a function of barrier transparency ratio (for arbitrary gap ratios $r_s$, $r_\Delta$) and as a function 
of temperature (for some values of the gap ratios). Here, the obtained current-phase relation was purely sinusoidal 
irrespective of parameter values, a result which can be explained by the fact that the linear Usadel equation corresponds 
to only a 1st order approximation in the interface resistance, so that no 2nd harmonic terms will appear. Our present results, 
on the other hand, are valid for arbitrary interface resistance, and we see that in this model the 2nd harmonic term plays a 
crucial role in the behavior of the Josephson junction, which we will discuss more below. We should also remark here on 
the difference between the diffusive and the ballistic model in that the former in contrast to the latter has a interlayer 
with finite thickness, which was needed to
justify the assumption of weak proximity effect.

The importance of a prospective 2nd harmonic contribution to the Josephson current is natural when we are concerned with 0-$\pi$ 
transitions, as this component may dominate when the 1st harmonic component vanishes close to the transition point. This fact, 
and the $\varphi$-junction behavior that follows, has been pointed out several times in the context of S$|$F$|$S junctions
\cite{golubov_cpr_rmp, konschelle-buzdin_nonsin_strong_fm, mohammadkhani-zareyan_sfs_2nd_harmonic}. For our model, the influence 
of the 2nd harmonic is seen to be particularly prevalent in the case of different gap magnitudes and/or high interface 
transparency. Before discussing its implication on the thermal phase shift effect observed here, we consider the case of 
more conventional 0-$\pi$ transitions. Most often when establishing 0-$\pi$ transitions in a Josephson junction, one looks 
for a sharp cusp in the critical current as a function of the parameter in question, although this method cannot discern 
which side of the transition represents the 0-state and which represents the $\pi$-state. By using a rf SQUID configuration
\cite{golubov_cpr_rmp} one may however measure the jump in the critical phase difference across the junction, which for a 
sinusoidal current-phase relation would be from $\varphi^\ast = \pi/2$ to $\varphi^\ast = -\pi/2$ or vice versa. (Note that 
it is crucial to this argument that one considers a current-biased experiment in which a current $I > 0$ is forced through 
the junction, letting the phase difference adjust accordingly.) In the presence of higher harmonics in the current-phase 
relation the jump from $\varphi^\ast > 0$ to $\varphi^\ast < 0$ or vice versa will in general be different, \cf the 
transition for varying $r_Z$, but the principle remains the same. This also holds when the sinusoidal component to the 
current-phase relation is subdominant for all parameter values, such as for the thermal transitions reported here, so 
that the critical current is not even close to zero at the transition point. Inspecting Fig. \ref{fig:Ic_vs_T_rS05_rD03_w_phi}, 
one sees that the critical current does in fact reach a minimum at $T = T_\varphi$. Hence this phase-shift effect can be 
regarded as a degenerate form of 0-$\pi$ transition which can only be established by SQUID measurement of the critical 
phase difference.
Alternatively, one could of course demonstrate the transition by using SQUID to map out the entire current-phase relation\cite{frolov-ryazanov_cpr}, but observing a single phase-shift of the critical phase $\varphi^\ast$ may 
be simpler experimentally. 

Considering then the peak phenomenon described for the temperature dependence of the critical current, as pointed out earlier, it does 
not pertain to the $s_\pm$-state per se, but is a general result in this framework of two gaps crossing at a certain temperature. In fact, this even holds when none of the two superconductors are multiband superconductors. 
Experimentally, this can however be understood to be a somewhat artificial situation, as the phenomenon would not occur for a junction consisting of two conventional 
superconductors with different zero-temperature gap magnitudes because of the universal ratio $2\Delta(0)/T_c = 1.76$ for 
BCS superconductors. And for \eg high-$T_c$ cuprates, for which the corresponding ratio is larger, 
the value of $\Delta(0)$ is typically much larger than for any conventional superconductor as well. So although multiband 
superconductors are not necessary as such, the described situation can occur here much more easily because the superconducting pairing for both bands typically vanish at the same critical temperature, whereas the gap ratio $r_\Delta \neq 1$. This is the 
situation for the conventional multiband superconductor MgB$_2$\cite{Brinkman_MgB2}, and also seems to be the case for the iron-based
 superconductors\cite{gonnelli_2_nodeless_gaps}. We should note that similar behavior was not found in the diffusive 
 case\cite{linder_iron_rapid}, but that a finite temperature maximum in the critical current for multiband superconductors was predicted in Ref. ~\onlinecite{Agterberg_multiband}. 
In that case, the effect was however ascribed to thermal effects combined with different sign of the two order parameters, and is not related to gap crossing irrespective of the order parameter sign as in our case. Furthermore, in Ref. ~\onlinecite{Agterberg_multiband} as well as in our results for the diffusive case, the current-phase relation was implicitly assumed to be purely sinusoidal, which may explain some of the differences with our present results for the ballistic case.

\begin{figure}[h]
	\centering
	\resizebox{0.35\textwidth}{!}{
	\includegraphics{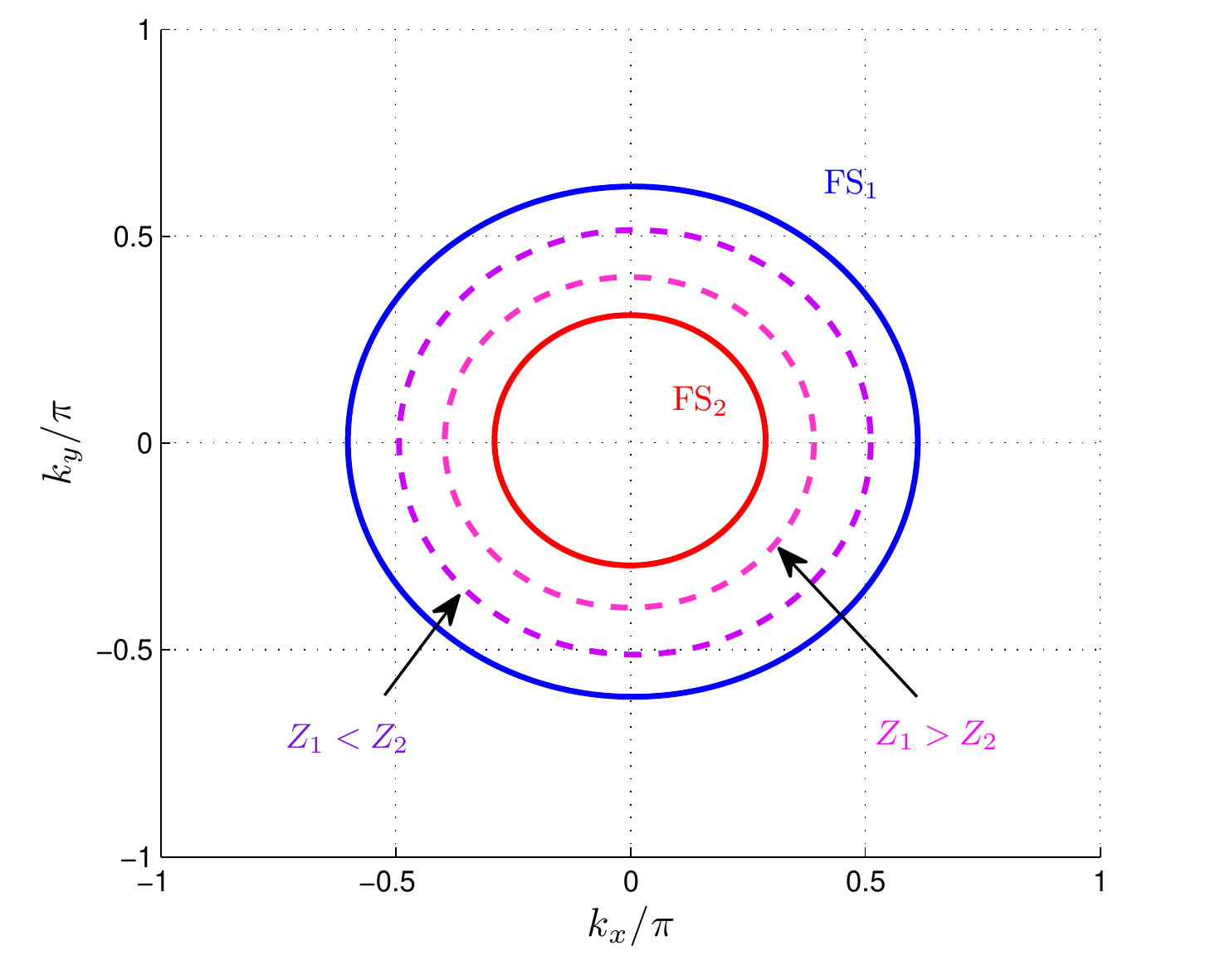}
	} \caption{(Color online) Illustration of the physical interpretation of the barrier strength ratio $r_Z$. The outer circle 
	represents the Fermi surface of band 1 of the $s_\pm$-wave superconductor and the inner circle represents band 2. The two 
	dashed circles in between represent the Fermi surface of the $s$-wave superconductor on the left hand side of the junction 
	for two situations: largest FWVM with respect to band 1 ($Z_1 > Z_2$) and largest FWVM with respect to band 2 ($Z_1 < Z_2$).}
	\label{fig:FS}
\end{figure}

As the dependence of various observable quantities on the barrier strength ratio $r_Z$ was considered frequently throughout 
Sec. \ref{sec:jos}, we would now like to present a more thorough rationale for this parameterization. Firstly, we note that 
although our model assumes the same Fermi wavevector $k_{\rm F}$ for all bands in all regions of our setups, any FWVM between 
the different regions is equivalent with an increase in the barrier strength $Z$. And for different Fermi wavevectors 
$k_{\mathrm{F},s}$, $k_{\mathrm{F},\lambda}$ for the $s$-wave superconductor on the left hand side and band 1,2 of the 
$s_\pm$-wave superconductor on the right hand side in the Josephson junction, respectively, this gives rise to different 
effective barrier strengths $Z_{1,2}$ for the two bands. This line of thought is illustrated in Fig. \ref{fig:FS}. The 
idea of tailoring the experimental setup by the use of materials with appropriate Fermi surfaces was first proposed in Ref.~\onlinecite{tsai-bernevig_cc}, and as we discussed in Ref. ~\onlinecite{linder_iron_rapid}, it might be possible 
to produce a series of junction samples with different barrier strength ratios $r_Z$ by varying the doping level in 
the non-$s_\pm$-wave region of the junction. In this manner, it is conceivable that a 0-$\pi$ transition can be 
observed for varying doping level, analogously as to how 0-$\pi$ transitions are observed S$|$F$|$S junctions for 
varying interlayer thickness. The preceding argument naturally raises the question whether it might be more appropriate 
with a parameterization in which increased transmittance through the barrier for one of the bands was accompanied by 
decreased transmittance for the other band. We did nevertheless define $r_Z$ as simply the effective barrier strength 
ratio because it is hard to tell exactly how the relative transmittance will change with doping level. This does 
probably also make it quite challenging to experimentally produce the right series of samples to observe 0-$\pi$-transitions, 
which is what makes possible observable signatures for varying temperature all the more appealing.

Finally, we discuss our model in context of the recently discovered iron-based superconductors. It should be stressed 
that our model is to be taken as a minimal model describing the generic behavior of transport phenomena in a two-band 
$s_\pm$-superconductor, 
but we would like to point out how a more realistic model should take into account the specifics of the iron-based superconductors.
Firstly, the BTK approach does not incorporate any details of the band structure, and spherical Fermi surfaces are assumed. 
Secondly, ours is a two-band model whereas it has been argued that at least four bands, two hole-like ($h$) and two 
electron-like ($e$), should be included to capture the physics behind the superconductivity in these 
materials\cite{cvetkovic-tesanovic_multiband}. The main effect of their inclusion from the point of view of transport 
would be the possibility of $e$-$e$ and $h$-$h$ interband scattering between nearly-degenerate electron and hole bands, 
respectively. Since these scattering processes would involve no internal phase-shift, we expect that the result would 
be qualitatively similar to the two-band case. This assumption can also be justified by the fact that $h$-$h$ and 
$e$-$e$ scattering processes should be weak in iron-based superconductors compared to the spin density-wave-enhanced 
$e$-$h$ interband processes\cite{Cvetkovic-Tesanovic_Valley_density-wave}, so that we to a good approximation can 
consider degenerate $e$ and $h$ bands. Generalization to a non-degenerate four-band model could nevertheless be made 
in our theory by a straightforward extension of Eqs. \eqref{eq:ham} and \eqref{eq:basis}, with the inclusion of 
$h$-$h$ and $e$-$e$ interface scattering terms in Eq. \eqref{eq:hop}, although an analytical treatment in that 
case would be a daunting task. Furthermore, one might have gap magnitudes that were momentum dependent, but the 
approximation of constant $s$-wave gaps on each of the Fermi surfaces should be reasonable. (The possibility of a 
$d$-wave gap or other pairing symmetries with nodes on the Fermi surface is left out of the question in this work, 
since the majority of experiments so far seem to indicate a nodeless gap on the Fermi surface.) Another extension 
would be to include interband scattering in the bulk of the $s_\pm$-superconductor, and not only near the interfaces 
as in our case, or even more sophisticated models \eg including momentum dependence in $\alpha$.

Regarding the magnetic field dependence of the critical Josephson current described in Sec. \ref{sec:fraunhofer}, we may 
compare our results with the experimental results for iron-based superconductors available at the moment. Inspecting the 
diffraction pattern in Fig. 3 of Ref. ~\onlinecite{zhang_iron_josephson}, we note an intriguing similarity with ours 
for $r_Z \gtrsim 1$ in that the critical current is nonvanishing between the diffraction maxima. This may however 
just as well be the combined result of nonuniform current distribution, trapped flux and deviation from the 
small junction limit\cite{vanharlingen_cuprates_rmp},
so that we cannot with any certainty interpret this observation as evidence for a non-sinusoidal current-phase relation, 
nor would non-sinusoidality necessarily imply $s_\pm$-wave pairing. (The diffraction patterns obtained in 
Refs. ~\onlinecite{zhou_iron_fraunhofer} and ~\onlinecite{zhang_iron_josephson2} can on the other hand not 
be compared with our results at all, as the experimental situations for those works are different.) It should 
also be noted that our modelling of the 
flux threading the junction is rather simplified, and does not include effects that may be present in real samples
\cite{vanharlingen_cuprates_rmp}.
More importantly, assuming isotropic order parameters and Fermi surfaces, our model is insensitive to how the junction 
geometry is chosen. We therefore cannot capture the directionality of the electron-like Fermi surfaces in the folded 
Brillouin zone of iron-based superconductors, which is essential in other proposals for phase-sensitive corner 
junctions \cite{wu_phillips_iron_squid, parker-mazin_phase_sensitive} and related geometries.

It would also be very interesting to see how robust the results presented here are to the introduction of material 
impurities. The iron-based superconductors are mostly expected to reside in some intermediate regime of impurity 
concentration\cite{mazin_review}, thereby making neither 
the ballistic nor the diffusive limit a completely accurate description. In fact, a number of theoretical
 works\cite{tsai_cc-pairing_impurity-induced_ABS,zhou_cc-pairing_impurity-induced_abs, matsumoto_impurity_bound_state, ng_impurity_bound_states,bang_spm_impurities,dolgov-golubov-parker} depend upon a significant influence of 
impurities to explain the experimental results or to induce experimentally observable bound states. Our 
study in Ref. ~\onlinecite{linder_iron_rapid} was motivated by the
fact that the diffusive regime is often the experimentally relevant one. Although taking the diffusive limit may 
not be strictly valid in this case, the results found might nevertheless capture important features of 
the real materials. In light of this, it would be very interesting to compare the results obtained in the 
diffusive and the ballistic limit with calculations performed using the quasiclassical Eilenberger 
equation\cite{eilenberger}, which allows for arbitrary impurity concentration. This would require a 
multiband extension of the Zaitsev boundary conditions\cite{zaitsev}, and such a theory has only very 
recently been developed (see Ref. ~\onlinecite{eschrig_boundary_cond}).

\section{Conclusion}\label{sec:concl}

Possible signatures of $s_\pm$-wave pairing in tunneling spectroscopy stem mainly from the multigap nature of the superconductor, but also 
from interference effects when the interband coupling is strong relative to the barrier strength. This may lead to subgap peaks in the 
conductance spectra not present for a corresponding $s$-wave model, although the appearance of these are relatively sensitive to the 
parameter values used. Similarly for the nonlocal conductance, it is found to be very difficult to discriminate qualitatively the 
interference effects of a $s_\pm$-wave state from those of a two-band $s$-wave state. 
Josephson coupling is on the other hand an intrinsically phase-dependent phenomenon, so it is natural that it is here that we find the 
most promising signatures of $s_\pm$-wave pairing, namely 0-$\pi$ transitions or similar phase shifts in a $s$-wave$|$I$|s_\pm$-wave junction. 
These are neither dependent on, nor considerably affected by, the presence of interband coupling. As in the diffusive case\cite{linder_iron_rapid}, 
we find 0-$\pi$ phase shifts as a function of the relative interface transparency, an effect whose detection is possible in principle, 
but difficult in practice. We have also shown that a phase-shift effect is present as a function of temperature, and although this effect 
is not as robust as the one reported for the diffusive case, it may nevertheless be possible to observe using a SQUID setup. For both cases, 
we have shown how the phase shifts can be ascribed to the competition between Andreev bound states for the two bands, and how the 
non-sinusoidality of the Josephson current is essential in the description of the phase shifts. We have also pointed out that this 
2nd harmonic component in the current-phase relation may induce half-integer quantum flux modulations in the magnetic diffraction 
pattern of the Josephson junction. In addition, we found a peak feature in the temperature dependence of the critical current 
for the case of different gap magnitudes, an effect ascribed to the crossing of two gaps. Although it is hard to tell how relevant 
the signatures reported in this simplified model are for possible experimental realizations of the $s_\pm$-wave pairing state, 
our results shed  more light on the basic mechanisms of transport and their implications in such systems.

\acknowledgments
J. L. and A. S. were supported by the Research Council of Norway, Grants No. 158518/431 and No. 158547/431 (NANOMAT), 
and Grant No. 167498/V30 (STORFORSK). A. S. thanks A. Balatsky and Z. Tesanovic for discussions, and acknowledges
the hospitality of the Aspen Center for Physics.  

\appendix

\section*{Appendix A: Reflection and transmission coefficients for the N$|s_\pm$-wave junction}
\noindent

In this section, we give the analytical solution for the reflection and transmission coefficients.  We have to consider 
the two cases $\lambda' = 1, 2$ for the incoming electron band independently, but will use the same symbols for the 
coefficients to simplify notation. First considering $\lambda' = 1$, we have the transmission coefficients given by
\begin{align}
	s_1 = \frac{2 \i R_2}{\Gamma}, \\
	t_1 = \frac{-2 \i R_1}{\Gamma}, \\
	s_2 = \frac{2 \i \talpha}{\Gamma} \frac{ X_{11} R_2 - X_{12} R_1 }{\gamma_2}, \\
	t_2 = \frac{2 \i \talpha}{\Gamma} \frac{ X_{21} R_2 - X_{22} R_1 }{\gamma_2}.
\end{align}
For the case of $\lambda' = 2$, the corresponding expressions read
\begin{align}
	s_1 = \frac{2 \i \talpha}{\Gamma} \frac{ X_{11} P_2 - X_{12} P_1 }{\gamma_1}, \\
	t_1 = \frac{2 \i \talpha}{\Gamma} \frac{ X_{21} P_2 - X_{22} P_1 }{\gamma_1}, \\
	s_2 = \frac{2 \i P_2}{\Gamma} , \\
	t_2 = \frac{-2 \i P_1}{\Gamma} . 
\end{align}
The reflection coefficients are then found for both cases by insertion into
\begin{align}
	r_1 & = -\delta_{\lambda',1} + u_1 s_1 + v_1 t_1,\\
	r_1^\text{A} &= v_1 s_1 + u_1 t_1,\\
	r_2 &= -\delta_{\lambda',2} + u_2 s_2 + \delta v_2 t_2,\\
	r_2^\text{A} &= \delta v_2 s_2 + u_2 t_2,
\end{align}
where $\delta_{\lambda',i}$ is the Kronecker delta.

The auxiliary quantities used for $\lambda' = 1, 2$ are given by
\begin{align}
	\Gamma &= \gamma_2 \gamma_1 + 2 \talpha^2 (4 u_1 u_2 A - Z^2 C_2 C_1) + \talpha^4 C_1 C_2,\\
	X_{11} & = Z A + 2 \i u_1 u_2,\\
	X_{22} &= Z A - 2 \i u_1 u_2,\\
	X_{12} &= Z B + 2 \i u_2 v_1,\\
	X_{21} &= Z B - 2 \i u_2 v_1,\\
	Y_{12} &= -Z B + 2 \delta \i u_1 v_2,\\
	Y_{21} &= -Z B - 2 \delta \i u_1 v_2,\\
	R_1 &= - Z v_1 \gamma_2 + \talpha^2(\delta v_2 X_{11} + u_2 X_{21}),\\
	R_2 &= -(2 \i + Z) u_1 \gamma_2 + \talpha^2(\delta v_2 X_{12} + u_2 X_{22}),\\
	P_1 &= -\delta Z v_2 \gamma_1 + \talpha^2(v_1 X_{11} + u_1 Y_{21}),\\
	P_2 &= -(2 \i + Z) u_2 \gamma_1 + \talpha^2(v_1 Y_{12} + u_1 X_{22}),
\end{align}
where
\begin{align}
	A &= u_1 u_2 - \delta v_1 v_2,\\
	B &= v_1 u_2 - \delta u_1 v_2,\\
	C_\lambda &= v_\lambda^2 - u_\lambda^2,\\
	\gamma_\lambda &= 4 u_\lambda^2 - C_\lambda Z^2.
\end{align}
The expressions above are valid both the $s_\pm$-wave and the coupled $s$-wave case, where $s_\pm$-wave is found 
by setting $\delta = -1$ and $s$-wave by $\delta = 1$.

\section*{Appendix B: Solution for the ABS energies for different gap magnitudes}
\noindent

The coefficient matrix $\Lambda$ for each of the uncoupled bands in the general case of different gap magnitudes 
yield after some manipulation the equation
\begin{widetext}
\begin{equation}
	\label{eq:imag_det_lambda}
	\Imag{|\Lambda_\lambda|} = (4+Z_\lambda^2) \sin{(2\beta_s + 2\beta_\lambda)} - Z_\lambda^2 \left[ \sin{2\beta_s} + \sin{2 \beta_\lambda} \right] - 8 \delta_\lambda \sin{(\beta_s + \beta_\lambda)} \cos{\varphi} = 0,
\end{equation}
with $\delta_1 = 1$ and $\delta_2 = - 1$ for a $s_\pm$-wave superconductor. Using that $\cos{\beta_\lambda} = E/|\Delta_\lambda|$ and $\cos{\beta_s} = |\Delta_\lambda|/\Delta_s \cos{\beta_\lambda}$ we can solve the equation for $\cos^2{\beta_\lambda}$, which produces the solutions
\begin{align}
	\label{eq:diffgaps1}
	E_1^\pm = \pm \frac{Z_1^2 + 2}{Z_1} \sqrt{ \frac{ A_1 \sin^2{(\varphi/2)} + B_1 - \sqrt{ C_1 \sin^4{(\varphi/2)} + D_1 \sin^2{(\varphi/2)} + F_1 } } { 2(Z_1^2 + 4 )} }, \\
	\label{eq:diffgaps2}	
	E_2^\pm = \pm \frac{Z_2^2 + 2}{Z_2} \sqrt{ \frac{ A_2 \cos^2{(\varphi/2)} + B_2 - \sqrt{ C_2 \cos^4{(\varphi/2)} + D_2 \cos^2{(\varphi/2)} + F_2 } } { 2(Z_2^2 + 4 )} },
\end{align}
\end{widetext}
in addition to several other unphysical solutions. The auxiliary quantities here are given by
\begin{align}
	A_\lambda = 2 K_\lambda, \\
	B_\lambda = \Delta_s^2 + |\Delta_\lambda|^2 - K_\lambda,\\
	C_\lambda = 8 K_\lambda \Delta_s |\Delta_\lambda|,\\
	D_\lambda = 4(\Delta_s - |\Delta_\lambda|)^2 K_\lambda,\\
	F_\lambda = (\Delta_s - \Delta_\lambda^2)^2,\\
	K_\lambda = 8 \Delta_s |\Delta_\lambda| / (Z_\lambda^2 + 2)^2.
\end{align}

To justify that the given solutions are the only solutions and are also in fact valid for all parameters, we have verified numerically 
that $\Real{|\Lambda_\lambda|} = \Imag{|\Lambda_\lambda|} = 0$ for all solutions of $E_\lambda^\pm$ used in this work. However, as 
can be seen by comparing with Fig. \ref{fig:ABS_rS05_rD03_T0855-0875} and the accompanying discussion, evaluating $E_\lambda(\varphi)$ 
for around $\varphi \approx 0$ for Eq. \eqref{eq:diffgaps1} or around $\varphi \approx \pm \pi$ for  Eq. \eqref{eq:diffgaps2} does 
not produce a valid result for $|\Delta_\lambda| \neq \Delta_s$. The explanation is that the physical Andreev bound states simply 
vanish in these regions, and we have again confirmed numerically that $|\Lambda_\lambda| = 0$ have no real solution for $E$ here. 
In fact, solving only for the imaginary part of the determinant yields (clearly unphysical) solutions 
$|E_\lambda| > \min \{\Delta_s, |\Delta_\lambda|\}$, which furthermore result in complex factors $\sin{\beta_\lambda}$, rendering 
Eq. \eqref{eq:imag_det_lambda} invalid as an expression for the imaginary part of the determinant. In the results presented 
above, we have handled this numerically by setting the bound state energy equal to the gap value when vanishing, so that 
it does not contribute to the current (since the energy states vanish at the gap edge with zero slope), although the energy 
states strictly speaking do not exist at all in these regions.


\end{document}